 \documentclass[10pt,pra,aps,a4paper,oneside,twocolumn,superscriptaddress,showpacs]{revtex4-2}

\usepackage{enumitem}

\usepackage[english]{babel}
\usepackage[latin1]{inputenc}
\usepackage[T1]{fontenc}
\usepackage{lmodern}
\usepackage{ulem}
\usepackage{dcolumn}
\usepackage{subfigure}
\usepackage{footnote}
\usepackage{float}
\usepackage{multirow}
\usepackage{amsmath,amsfonts,amssymb}

\usepackage{graphicx}  
\usepackage{latexsym}   
\usepackage{color}

\begin{document}

\title{Electric and magnetic toroidal dipole excitations in core-shell spheres}

\author{\firstname{Tiago}  Jos\'e \surname{Arruda}}
\email{tiago.arruda@unifal-mg.edu.br}
\affiliation{Instituto de Ci\^encias Exatas (ICEx),
Universidade Federal de Alfenas (UNIFAL-MG),\\
 37133-840 Alfenas, Minas Gerais, Brazil}

\begin{abstract}

Toroidal dipole moments, which consist of enclosed circulating currents aligned along paths within a torus shape, can be experimentally achieved in metamaterials using various geometrical configurations. Here, we investigate the excitation of toroidal dipole moments in a core-shell nanosphere composed of dispersive materials within the framework of the Lorenz-Mie theory.
By isolating the contributions related to volumetric oscillations of electric charges from those of transverse and radial current oscillations, we derive closed-form analytic expressions for the scattering coefficients associated with electric and magnetic toroidal dipole moments.
These analytic expressions, obtained beyond the Rayleigh scattering regime and previously concealed within the Lorenz-Mie theory, agree with the Cartesian toroidal dipoles determined from the radial and transverse current oscillations.
By exploring the resonances in these calculated coefficients, we demonstrate that the electric and magnetic toroidal dipole excitations can be manipulated and tuned for near- and far-field applications involving plasmonic core-shell nanoparticles and engineered metamaterials.
We believe that our analytic results could shed light on some of the extraordinary effects obtained in Lorenz-Mie theory that depend on the interference of multipole coefficients, such as toroidal dipole-induced transparency, Fano resonances and superscattering of light.

\end{abstract}


\maketitle


\section{Introduction}

The scattering of electromagnetic waves by obstacles smaller than the incident wavelength is a classical topic that features in most of the recent advances in optics and nanophotonics.
Theoretically, this topic is commonly analysed using the multipole expansion technique~\cite{Grahn_NJPhys14_2012,Li_PhysRevB97_2018}.
In this approach, the electric and magnetic contributions to scattering are generally well described by the lowest-order multipoles, which are the electric and magnetic dipoles.
However, higher-order multipole contributions become relevant when the wavelength of the incident wave is large compared to the size of the scatterer, requiring consideration of additional terms in the multipole expansion~\cite{Grahn_NJPhys14_2012,Alaee_OptCommun2018}.
With the advent of engineered metamaterials, it has been demonstrated that specifically designed subwavelength structures can still excite and manipulate higher-order multipole moments.
This has led to several applications of intriguing optical effects such as Fano resonances~\cite{Luk_NatMater9_2010,Arruda_PhysRevA87_2013,Arruda_PhysRevA92_2015}, plasmonic fluorescence enhancement~\cite{Vandenbem_PhysRevB81_2010,Arruda_PhysRevA96_2017,Arruda_PhysRevB98_2018}, optical magnetism~\cite{Kallos_PhysRevB86_2012,Liu_Nano9_2020}, directional scattering~\cite{Luk_ACSPhot2_2015,Trigo_PhysRevLett125_2020}, plasmonic cloaking~\cite{Chen_AdvMater24_2012,Fleury_PhysRevAppl4_2015}, bound states in the continuum~\cite{Koshelev_PhysRevLett121_2018}, among others.

In recent years, there has been a growing interest in researching dynamic toroidal multipole moments for light scattering applications~\cite{Kaelberer_Science330_2010,Savinov_PhysRevB89_2014}.
These multipole excitations constitute a class of multipole moments obtained through decomposition of moment tensors, commonly referred to as the Cartesian multipole moments.
While the conventional electric multipoles arise from the separation of oscillating charges with opposite signs, toroidal multipoles are induced by the enclosed circulation of electric currents flowing on the surface of a torus along its meridians~\cite{Radescu_PhysRevE65_2002}.
Although the conventional multipole expansion in vector spherical harmonics is complete~\cite{Muhlig_Metamat5_2011}, it has been demonstrated that it conceals the presence of Cartesian toroidal multipoles within the expressions, providing renormalized electric and magnetic multipoles (dipole, quadrupole, etc.)~\cite{Zhang_PhysRevA92_2015}.
This is the case with the classical Lorenz-Mie theory, which analytically describes the electromagnetic scattering of a plane wave by a single sphere~\cite{Bohren_Book_1983} and is widely exploited to manipulate electromagnetic fields in a variety of applications~\cite{Liu_Nano9_2020,Arruda_PhysRevA101_2020,Dorodnyy_LaserPhotRev17_2023}.
Within the framework of the Lorenz-Mie solution, the electric and magnetic contributions are renormalized, and the toroidal multipole moments are concealed within the well-known Lorenz-Mie scattering coefficients, $a_{\ell}$ and $b_{\ell}$~\cite{Kivshar_NatComm6_2015}.

Historically, the static toroidal dipole moment was first introduced in 1958 by Zel'dovich in the context of parity violation in nuclear physics~\cite{Zeldovich_JExpTheorPhys33_1957}.
Since then, the toroidal dipole moment has been referred to as an ``anapole'' as it has no correspondence with the usual electric and magnetic multipole moments~\cite{Kivshar_NatComm6_2015}.
Although the importance of toroidal dipoles has been widely recognized in solid-state and nuclear physics, as well as in classical electrodynamics~\cite{Dubovik_PhysRep187_1990,Radescu_PhysRevE65_2002}, it was only with the advent of the electromagnetic metamaterials that toroidal dipole excitations were demonstrated to play a crucial role in the areas of optics and photonics~\cite{Kaelberer_Science330_2010,Zheludev_NatMat15_2016,Talebi_Nanoph7_2018}.
This is mainly due to the fact that the far-field radiation of a dynamic toroidal dipole moment is indistinguishable from the electric and magnetic dipoles in the conventional multipole expansion of electromagnetic scattered fields.
However, the interplay between near- and far-field radiation in optical structures of size comparable to the wavelength has been shown to be affected by the dynamic toroidal dipole response.
Indeed, dynamic toroidal dipole excitations have been observed in plasmonic~\cite{Huang_OptExpress20_2012,Ogut_NanoLett12_2012,Ge_OptExpress25_2017} and dielectric~\cite{Basharin_PhysRevX5_2015,Ospanova_OptLett43_2018,He_PhysRevB98_2018} metamaterials across frequencies ranging from microwave to near-infrared and visible frequency ranges.
This unusual class of multipole moments could have potential applications in nano-lasers~\cite{Huang_SciRep3_2013}, sensors~\cite{Ye_PhysScr88_2013,Gupta_ApplPhysLett110_2017,Ahmadivand_ACSSens2_2017,Wang_Nanoph10_2021}, optical radiation manipulation~\cite{Zhou_OpticsExp28_2020}, photoluminescence emission~\cite{Cui_Nanoscale11_2019}, harmonic generation~\cite{Ahmadivand_NanoLett19_2018}, data storage~\cite{Spaldin_JPhysCondensMatter20_2008}, and so on.

Recently, the excitation of toroidal multipoles within single scattering spheres has been demonstrated to play a fundamental role in some exotic scattering phenomena, such as non-radiating anapole modes in dielectric nanoparticles~\cite{Kivshar_NatComm6_2015}, toroidal dipole-induced transparency~\cite{Miroshnichenko_LaserPhotRev9_2015} and toroidal dipole-induced absorption~\cite{Jiang_OptExpress25_2017} in core-shell nanospheres.
Since dynamic toroidal multipoles are not explicitly accounted for in the Lorenz-Mie theory, all of these studies identify the toroidal dipole contributions via the Cartesian toroidal dipole expansion applied to the electric field distribution within the scatterer.
\color{black}
Specifically, one has to calculate the near-field distribution, which consists of cumbersome summations of vector spherical harmonics weighted by Lorenz-Mie coefficients, and then numerically integrate it over the scatterer's volume to analyze toroidal dipole excitations in plasmonic and dieletric spheres.
There are no analytic expressions in the Lorenz-Mie theory that provide directly the toroidal contributions to light scattering, which are hidden in the spherical multipoles.
\color{black}

In this paper, we precisely distinguish the excitation of electric and magnetic toroidal dipoles from the spherical (renormalized) electric and magnetic dipole moments calculated within the framework of the Lorenz-Mie theory.
We achieve this by studying analytically the electromagnetic fields inside a core-shell sphere, expanded in terms of vector spherical harmonics.
By considering the limiting case of a small-diameter sphere compared to the wavelength, we expand the spherical Bessel functions in powers of the size parameter and segregate the contributions related to volumetric oscillations of electric charges from the transverse and radial current oscillations.
Integrating each contribution over the scatterer's volume, we derive closed-form analytic expressions for the partial scattering coefficients associated with the dynamic electric and magnetic toroidal dipole moments.
These analytic expressions precisely replicate the Cartesian toroidal multipoles calculated from the integration over terms related to the radial and transverse current oscillations, and could serve as important benchmarks for numerical calculations in more complex geometries.
Through an analysis of the resonances in the derived coefficients, we demonstrate the potential for manipulating and tuning toroidal dipoles for near- and far-field applications involving plasmonic core-shell nanoparticles.

The remainder of this paper is organized as follows.
In Sec.~\ref{multipole-revision}, we revisit the main expressions concerning the multipole expansion method and the scattering coefficients in spherical coordinates, and define the coefficients related to electric and magnetic toroidal multipoles.
Section~\ref{Cartesian-multipoles} briefly presents the main expressions commonly used to calculate the electric and magnetic toroidal dipole moments in the Cartesian multipole expansion method.
The main analytic results of this paper are in Sec.~\ref{Lorenz-Mie}, where we derive closed analytic expressions for scattering coefficients related to electric and magnetic toroidal dipole moments within the framework of the Lorenz-Mie theory.
Section~\ref{Numerical} presents a discussion on light scattering by plasmonic core-shell nanoparticles containing a gain medium.
We demonstrate in Secs.~\ref{Ag-core} and \ref{Ag-shell} that both electric and magnetic toroidal dipole excitations in light scattering by core-shell spheres can be enhanced by a gain-assisted dielectric medium within the plasmonic scatterer.
Finally, in Sec.~\ref{conclusion} we summarize our main results and conclude.

\section{Theoretical foundation}
\label{theory}

\subsection{Multipole expansion and scattering coefficients}
\label{multipole-revision}

Let us consider the scattering of an electromagnetic plane wave by a finite obstacle embedded in a lossless homogeneous dielectric medium with permittivity $\varepsilon_0$ and permeability $\mu_0$.
Using spherical coordinates, the multipole expansion of a monochromatic electric field scattered by the obstacle, with amplitude $E_0$ and time harmonic dependence $e^{-\imath\omega t}$, can be expressed as~\cite{Grahn_NJPhys14_2012,Jackson_Book}
\begin{align}
\mathbf{E}_{\rm sca}(\mathbf{r}) &= E_0 \sum_{\ell = 1}^{\infty}\sum_{m=-\ell}^{\ell}\imath^{\ell}\sqrt{\pi(2\ell+1)}\nonumber\\
&\bigg\{a_{\rm E}(\ell,m)\frac{1}{k}\nabla\times\left[h_{\ell}^{(1)}(kr)\mathbf{X}_{\ell m}(\theta,\varphi)\right]\nonumber\\
&+a_{\rm M}(\ell,m)h_{\ell}^{(1)}(kr)\mathbf{X}_{\ell m}(\theta,\varphi)\bigg\},\label{E_sca}
\end{align}
where $h_{\ell}^{(1)}$ is the Hankel function of the first kind, $\mathbf{X}_{\ell m}(\theta,\varphi)=\mathbf{L}Y_{\ell m}(\theta,\varphi)/\sqrt{\ell(\ell+1)}$ is the (normalized) vector spherical harmonics, with $\mathbf{L}=-\imath\mathbf{r}\times\boldsymbol{\nabla}$ the angular momentum operator, $Y_{\ell m}$ is the spherical harmonics, and $k=\omega\sqrt{\varepsilon_0\mu_0}$ is the wave number \color{black} in the surrounding medium\color{black}.
The corresponding magnetic field $\mathbf{H}_{\rm sca}(\mathbf{r})$ is readily obtained by the curl's Maxwell equation: $\mathbf{H}_{\rm sca}(\mathbf{r})=-\imath \boldsymbol{\nabla}\times\mathbf{E}_{\rm sca}(\mathbf{r})/k\eta$, where $\eta=\sqrt{\mu_0/\varepsilon_0}$ is the  impedance of the surrounding medium.

The spherical multipole coefficients $a_{\rm E}(\ell,m)$ and $a_{\rm M}(\ell,m)$ in Eq.~(\ref{E_sca}) depend on the boundary conditions of the system.
The general expressions for these coefficients calculated by Grahn {\it et al.}~\cite{Grahn_NJPhys14_2012} are, respectively,
\begin{align}
a_{\rm E}(\ell,m)
&=\frac{(-\imath)^{\ell-1}k\eta}{E_0\sqrt{\pi(2\ell+1)\ell(\ell+1)}}\int{\rm d}^3rY_{\ell m}^*(\theta,\varphi)j_{\ell}(kr)\nonumber\\
&\left\{k^2{\mathbf{r}}\cdot\mathbf{J}(\mathbf{r})+\left(2+r\frac{{\rm d}}{{\rm d}r}\right)\left[\boldsymbol{\nabla}\cdot\mathbf{J}(\mathbf{r})\right]\right\},\label{a_E}\\
a_{\rm M}(\ell,m)
&=\frac{(-\imath)^{\ell-1}k^2\eta}{E_0\sqrt{\pi(2\ell+1)\ell(\ell+1)}}\int{\rm d}^3r Y_{\ell m}^*(\theta,\varphi)j_{\ell}(kr)\nonumber\\
&{\mathbf{r}}\cdot\left[\boldsymbol{\nabla}\times\mathbf{J}(\mathbf{r})\right],\label{a_M}
\end{align}
where $j_{\ell}$ is the spherical Bessel function and the source-current density is
\begin{align}
\mathbf{J}(\mathbf{r})=-\imath\omega\left[\varepsilon(\mathbf{r})-\varepsilon_0\right]\mathbf{E}(\mathbf{r}),\label{J}
\end{align}
with $\varepsilon(\mathbf{r})$ being the local electric permittivity associated with the scattering center and $\mathbf{E}(\mathbf{r})$ is the local electric field.

Equations~(\ref{a_E}) and (\ref{a_M}) describe the electric and magnetic properties of the interaction between an electromagnetic plane wave and a finite obstacle in free space.
In the case of an isotropic spherical scatterer \color{black} under plane wave incidence\color{black}, only the multipole moments with $m=1$ contribute to the sum in Eq.~(\ref{E_sca}), from which one retrieves the well-known Lorenz-Mie coefficients~\cite{Bohren_Book_1983}.
Using the notation of the classical Bohren and Huffman's book~\cite{Bohren_Book_1983}, the correspondence between Eqs.~(\ref{a_E}) and (\ref{a_M}) and the  Lorenz-Mie scattering coefficients is
\begin{align}
a_{\ell}=a_{\rm E}(\ell,1)\ ,\quad b_{\ell}=a_{\rm M}(\ell,1).
\end{align}
\color{black}
These coefficients are explicitly calculated in Appendix~\ref{an-and-bn}.\color{black}

Each term in Eqs.~(\ref{a_E}) and (\ref{a_M}) is related to different modes of localized charge distributions and current oscillations, which give rise to multipole moments~\cite{Zheludev_NatMat15_2016}.
Indeed, by the continuity equation, terms related to $\boldsymbol{\nabla}\cdot\mathbf{J}=\imath\omega \rho$, where $\rho$ is the electric charge density, correspond to volumetric oscillations of electric charges, which yield electric multipole moments.
Conversely, terms related to transverse $(\mathbf{r}\times\mathbf{J})$ or radial $(\mathbf{r}\cdot\mathbf{J})$ current oscillations yield magnetic or toroidal multipole moments, respectively.
Hence, one could define the coefficients $a_{\rm E}^{\rm c}(\ell,m)\propto\int {\rm d}^3r \rho Y_{\ell,m}^*\psi_{\ell}'(kr)$ for Cartesian electric contributions, $a_{\rm M}(\ell,m)\propto\int {\rm d}^3r (\boldsymbol{\nabla}\cdot[\mathbf{r}\times\mathbf{J}])Y_{\ell,m}^*j_{\ell}(kr)$ for magnetic contributions, and $a_{\rm E}^{\rm t}(\ell,m)\propto\int {\rm d}^3r (\mathbf{r}\cdot\mathbf{J})Y_{\ell,m}^*j_{\ell}(kr)$ for electric toroidal contributions, where $a_{\rm E}(\ell,m)= a_{\rm E}^{\rm c}(\ell,m)+a_{\rm E}^{\rm t}(\ell,m)$; see Box 2 in Ref.~\cite{Zheludev_NatMat15_2016}.
However, note that the term $(2+r\ {d}/dr)\boldsymbol{\nabla}\cdot\mathbf{J}(\mathbf{r})$ in Eq.~(\ref{a_E}) includes a radial derivative of the electric charge density and can also exhibits radial current oscillations.
This complicates the separation into Cartesian electric dipole and electric toroidal dipole contributions, making it not as straightforward as sometimes mentioned.

The correct separation between conventional multipoles and toroidal multipoles in spherical coordinates must take into account the dependence on the size parameter $kr$, similar to the Cartesian expansion of the multipoles.
To address this issue, one must first eliminate the spatial derivatives of $\mathbf{J}$ by integration by parts~\cite{Grahn_NJPhys14_2012} in Eqs.~(\ref{a_E}) and (\ref{a_M}):
\begin{align}
a_{\rm E}(\ell,m)&=\frac{(-\imath)^{\ell-1}k^2\eta F_{\ell m}}{E_0\sqrt{\pi(2\ell+1)}}\int{\rm d}^3r e^{-\imath m\varphi}\nonumber\\
&\bigg\{\left[\psi_{\ell}(kr)+\psi_{\ell}''(kr)\right]P_{\ell}^m(\cos\theta)\hat{\mathbf{r}}\cdot\mathbf{J}(\mathbf{r})\nonumber\\
&+\frac{\psi_{\ell}'(kr)}{kr}\left[\tau_{\ell m}\hat{\boldsymbol{\theta}}\cdot\mathbf{J}(\mathbf{r})-\imath\pi_{\ell m}\hat{\boldsymbol{\varphi}}\cdot\mathbf{J}(\mathbf{r})\right]\bigg\},\label{a_E2}\\
a_{\rm M}(\ell,m)&=\frac{(-\imath)^{\ell+1}k^2\eta F_{\ell m}}{E_0\sqrt{\pi(2\ell+1)}}\int{\rm d}^3r e^{-\imath m\varphi}\nonumber\\
&j_{\ell}(kr)\left[\imath\pi_{\ell m}\hat{\boldsymbol{\theta}}\cdot\mathbf{J}(\mathbf{r})+\tau_{\ell m}\hat{\boldsymbol{\varphi}}\cdot\mathbf{J}(\mathbf{r})\right],\label{a_M2}
\end{align}
where the angular functions are $\tau_{\ell m}={\rm d} P_{\ell}^m(\cos\theta)/{\rm d}\theta$ and $\pi_{\ell m}=m P_{\ell}^m(\cos\theta)/\sin\theta$, with $P_{\ell}^m$ being the associated Legendre functions, $\psi_{\ell}$ is the Riccati-Bessel function, and
\begin{align}
F_{\ell m}=\frac{1}{\sqrt{\ell(\ell+1)}}\sqrt{\frac{2\ell+1}{4\pi}\frac{(\ell-m)!}{(\ell+m)!}}.
\end{align}
Here we implicitly assume that the local magnetic permeability $\mu(\mathbf{r})=\mu_0$, meaning that all the media involved are nonmagnetic.
By substituting the recursive relations of spherical Bessel functions~\cite{Abramovitz_1964}, $\psi_{\ell}(kr)+\psi_{\ell}''(kr)=\ell(\ell+1){j_{\ell}(kr)}/{kr}$ and ${\psi_{\ell}'(kr)}/{kr}=(\ell+1){j_{\ell}(kr)}/{kr}-j_{\ell+1}(kr)$, into Eq.~(\ref{a_E2}), we finally obtain the electric coefficient
\begin{align}
a_{\rm E}(\ell,m)&=\frac{(-\imath)^{\ell-1}k^2\eta F_{\ell m}}{E_0\sqrt{\pi(2\ell+1)}}\int{\rm d}^3r e^{-\imath m\varphi}\nonumber\\
&\Bigg\{\ell(\ell+1)\dfrac{j_{\ell}(kr)}{kr}P_{\ell}^m(\cos\theta)\hat{\mathbf{r}}\cdot \mathbf{J}(\mathbf{r})\nonumber\\
&+\left[(\ell+1)\dfrac{j_{\ell}(kr)}{kr}-j_{\ell+1}(kr)\right]\nonumber\\
&\left[\tau_{\ell m}\hat{\boldsymbol{{\theta}}}\cdot\mathbf{J}(\mathbf{r})-\imath\pi_{\ell m}\hat{\boldsymbol{\varphi}}\cdot\mathbf{J}(\mathbf{r})\right]\Bigg\},\label{a_E3}
\end{align}
which agrees with Ref.~\cite{Li_PhysRevB97_2018}.

The coefficient $a_{\rm E}(\ell,m)$ expressed in terms of $j_{\ell}(kr)$ is well-suited for distinguishing electric dipole and electric toroidal dipole contributions in light scattering.
This is because, in comparison to the Cartesian electric dipole, the amplitude of the Cartesian electric toroidal dipole scales with two additional orders of $kr$.
Hence, considering $kr\ll 1$ (\color{black}long-wavelength approximation\color{black}), one has~\cite{Abramovitz_1964}
\begin{align}
j_{\ell}(kr)\approx \dfrac{(kr)^{\ell}}{(2\ell+1)!!}-\dfrac{(kr)^{\ell+2}}{2(2\ell+3)!!},\label{j-n}
\end{align}
where the leading term of $j_{\ell}$ depending on $(kr)^{\ell}$ is associated with the electric dipole amplitude for $\ell=1$, and the second term, depending on $(kr)^{\ell+2}$, is related to the electric toroidal dipole amplitude.
By substituting Eq.~(\ref{j-n}) into Eq.(\ref{a_E3}), we can define the partial scattering coefficients

\begin{align}
a_{\rm E}^{\rm c}(\ell,m)&=\frac{(-\imath)^{\ell-1}k^2\eta F_{\ell m}}{E_0\sqrt{\pi(2\ell+1)}}\int{\rm d}^3r e^{-\imath m\varphi}\frac{\ell+1}{(2\ell+1)!!}(kr)^{\ell-1}\nonumber\\
\Bigg\{&\ell P_{\ell}^m(\cos\theta)\hat{\mathbf{r}}\cdot \mathbf{J}(\mathbf{r})  + \tau_{\ell m}\hat{\boldsymbol{{\theta}}}\cdot\mathbf{J}(\mathbf{r})-\imath\pi_{\ell m}\hat{\boldsymbol{\varphi}}\cdot\mathbf{J}(\mathbf{r})\Bigg\},\label{a_E-ap}\\
a_{\rm E}^{\rm t}(\ell,m)&=\frac{(-\imath)^{\ell-1}k^2\eta F_{\ell m}}{E_0\sqrt{\pi(2\ell+1)}}\int{\rm d}^3r e^{-\imath m\varphi}\nonumber\\
&\Bigg\{-\ell(\ell+1)\dfrac{(kr)^{\ell+1}}{2(2\ell+3)!!}P_{\ell}^m(\cos\theta)\hat{\mathbf{r}}\cdot \mathbf{J}(\mathbf{r})\nonumber\\
&-\left[(\ell+1)\dfrac{(kr)^{\ell+1}}{2(2\ell+3)!!}+j_{\ell+1}(kr)\right]\nonumber\\
&\left[\tau_{\ell m}\hat{\boldsymbol{{\theta}}}\cdot\mathbf{J}(\mathbf{r})-\imath\pi_{\ell m}\hat{\boldsymbol{\varphi}}\cdot\mathbf{J}(\mathbf{r})\right]\Bigg\},\label{a_Te-ap}
\end{align}
where $a_{\rm E}(\ell,m)= a_{\rm E}^{\rm c}(\ell,m)+a_{\rm E}^{\rm t}(\ell,m)$.
The same applies to the magnetic coefficient in Eq.~(\ref{a_M2}), which can be easily separated into conventional magnetic multipoles and toroidal magnetic components by using Eq.~(\ref{j-n}):

\begin{align}
a_{\rm M}^{\rm c}(\ell,m)&=\frac{(-\imath)^{\ell+1}k^2\eta F_{\ell m}}{E_0\sqrt{\pi(2\ell+1)}}\int{\rm d}^3r e^{-\imath m\varphi}\nonumber\\
&\dfrac{(kr)^{\ell}}{(2\ell+1)!!}\left[\imath\pi_{\ell m}\hat{\boldsymbol{\theta}}\cdot\mathbf{J}(\mathbf{r})+\tau_{\ell m}\hat{\boldsymbol{\varphi}}\cdot\mathbf{J}(\mathbf{r})\right],\label{a_M-ap}\\
a_{\rm M}^{\rm t}(\ell,m)&=-\frac{(-\imath)^{\ell+1}k^2\eta F_{\ell m}}{E_0\sqrt{\pi(2\ell+1)}}\int{\rm d}^3r e^{-\imath m\varphi}\nonumber\\
&\dfrac{(kr)^{\ell+2}}{2(2\ell+3)!!}\left[\imath\pi_{\ell m}\hat{\boldsymbol{\theta}}\cdot\mathbf{J}(\mathbf{r})+\tau_{\ell m}\hat{\boldsymbol{\varphi}}\cdot\mathbf{J}(\mathbf{r})\right],\label{a_Tm-ap}
\end{align}
where $a_{\rm M}(\ell,m)= a_{\rm M}^{\rm c}(\ell,m)+a_{\rm M}^{\rm t}(\ell,m)$.
\color{black}
We emphasize that the approximation in Eq.~(\ref{j-n}) is applied only in the terms with $j_{\ell}(kr)$ and there is no assumption regarding the charge-current distribution $\mathbf{J}(\mathbf{r})$.
\color{black}

\subsection{The Cartesian toroidal dipoles}
\label{Cartesian-multipoles}

\begin{figure}[htbp]
\includegraphics[width=.8\columnwidth]{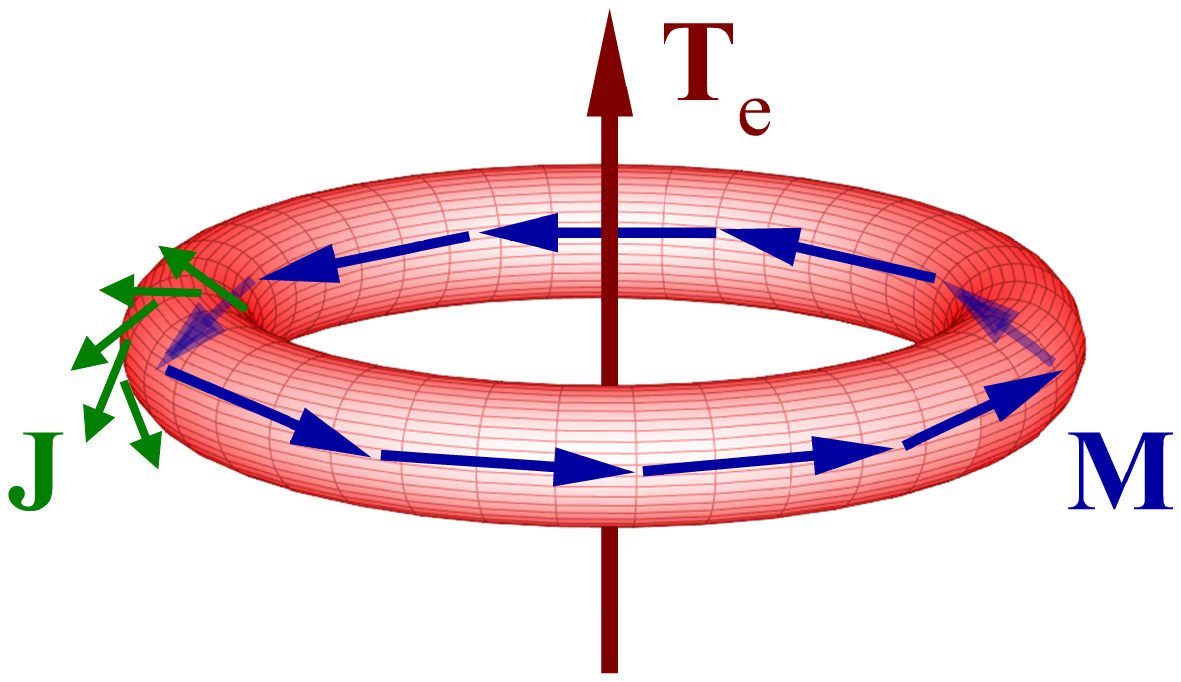}
\vspace{-1cm}
\caption{An electric toroidal dipole (ETD, $\mathbf{T}_{\rm e}$) is obtained by poloidal electric currents $\mathbf{J}$ flowing on the surface of a torus along its meridians or by a set of magnetic dipoles (MD, $\mathbf{M}$) arranged head-to-tail to form a closed loop.
Similarly, a closed loop of electric dipoles (ED, $\mathbf{P}$) gives rise to a magnetic toroidal dipole (MTD, $\mathbf{T}_{\rm m}$). }\label{fig1}
\end{figure}

In the electromagnetic wave scattering theory, results for the dipole approximation are obtained from the multipole expansion, considering scatterers with effective size $R$ much smaller than the wavelength $\lambda$ of the incident wave ($2\pi R/\lambda\ll1$).
This condition ensures that higher-order multipoles terms ($\ell>1$) can be disregarded in the multipole expansion.
In this case, the far-field radiation can be expressed as~\cite{Jackson_Book,Miroshnichenko_LaserPhotRev9_2015}
\begin{align}
\mathbf{E}_{\rm sca}(\bold{r})&\approx\frac{k^2 e^{\imath kr}}{4\pi r\varepsilon_0}\bigg[\hat{\mathbf{r}}\times(\mathbf{P}+\imath k\mathbf{T}_{\rm e})\times\hat{\mathbf{r}}\nonumber \\
&-\frac{1}{c}\hat{\mathbf{r}}\times\left(\mathbf{M}+k\mathbf{T}_{\rm m}\right)\bigg],\label{Erad}
\end{align}
where the electric $(\mathbf{P}$) and magnetic $(\mathbf{M})$ dipoles, and the electric $(\mathbf{T}_{\rm e})$ and magnetic $(\mathbf{T}_{\rm m})$ toroidal dipoles  are obtained, respectively, from the integrations of the charge-current distribution~\cite{Li_PhysRevB97_2018,Yang_Nanoscale12_2020}:
\begin{align}
\mathbf{P}&=\frac{\imath}{\omega}\int {\rm d}^3r \mathbf{J}(\mathbf{r}),\label{P}\\
\mathbf{M}&=\frac{1}{2}\int {\rm d}^3r \left[\mathbf{r}\times\mathbf{J}(\mathbf{r})\right],\label{M}\\
\mathbf{T}_{\rm e}&=\frac{1}{10c}\int {\rm d}^3 r\left\{\left[\mathbf{r}\cdot\mathbf{J}(\mathbf{r})\right]\mathbf{r}-2r^2\mathbf{J}(\mathbf{r})\right\},\label{Te}\\
\mathbf{T}_{\rm m}&=-\frac{k}{20}\int {\rm d}^3r r^2\left[\mathbf{r}\times\mathbf{J}(\mathbf{r})\right].\label{Tm}
\end{align}
\color{black}
Equation~(\ref{Erad}) is obtained from Chapter 10 of Ref.~\cite{Jackson_Book} by identifying the spherical dipoles $\mathbf{p}$ and $\mathbf{m}$ as the Cartesian dipoles $\mathbf{P}+\imath k\mathbf{T}_{\rm e}$ and $\mathbf{M}+k\mathbf{T}_{\rm m}$, respectively.
In Fig.~\ref{fig1} we show a schematic illustration of the electric toroidal dipole generated by the enclosed circulation of the magnetic dipole or by the electric current distribution in a toroidal configuration.
\color{black}

Equations~(\ref{P})--(\ref{Tm}) are generally referred to as the Cartesian electric and magnetic multipoles of first order and are widely used to calculate both electric and magnetic dipole and toroidal dipole contributions in light scattering.
The quantities $\mathbf{P}$, $\mathbf{M}$, $\mathbf{T}_{\rm e}$ and $\mathbf{T}_{\rm m}$ can be calculated from the local charge-current distribution defined in Eq.~(\ref{J}), which is usually associated with a scatterer that cannot be regarded as a perfect dipole in the far field~\cite{Miroshnichenko_OptExp23_2015}.
In particular, in the absence of the magnetic dipole, when $\mathbf{P}=-\imath k\mathbf{T}_{\rm e}$ one has the so-called ``anapole'', which corresponds to a configuration of non-vanishing and non-radiating charge-current distribution within the scatterer that leads to $\mathbf{E}_{\rm sca}=\mathbf{0}$~\cite{Miroshnichenko_LaserPhotRev9_2015}.
Similarly, from Eq.~(\ref{Erad}), the cancellation of the magnetic dipole contribution to the scattering occurs when $\mathbf{M}=- k\mathbf{T}_{\rm m}$, which is still a result not well explored in applications, as the magnetic dipole is usually accompanied by higher-order excitations, such as the electric quadrupole.
Although Eqs.~(\ref{P})--(\ref{Tm}) are valid only for far-field dipole radiation as expressed in Eq.~(\ref{Erad}), it is usual to consider them within the framework of multipole expansion.

The analysis of electromagnetic scattering using Cartesian multipoles typically involves numerically determining the scattered power associated with each multipole contribution.
The scattered powers linked to the Cartesian electric and magnetic dipoles are calculated, respectively, from~\cite{Miroshnichenko_LaserPhotRev9_2015}
\begin{align}
W_{\bf P}&=\dfrac{\mu_0\omega^4}{12\pi c}|\mathbf{P}|^2,\label{Wp}\\
W_{\bf M}&=\dfrac{\mu_0\omega^4}{12\pi c^3}|\mathbf{M}|^2,
\end{align}
whereas the power associated with the electric and magnetic toroidal dipoles are, respectively,
\begin{align}
\ W_{{\bf T}_{\rm e}}&=\dfrac{\mu_0\omega^4 k^2}{12\pi c}|\mathbf{T}_{\rm e}|^2,\\
\ W_{{\bf T}_{\rm m}}&=\dfrac{\mu_0\omega^4 k^2}{12\pi c^3}|\mathbf{T}_{\rm m}|^2.\label{Wtm}
\end{align}
Within the framework of the Lorenz-Mie theory, however, the toroidal multipole contributions to the scattered power are not distinguished from the conventional multipole contributions as in Eqs.~(\ref{Wp})--(\ref{Wtm}).
In fact, the spherical electric dipole and magnetic dipole $(\ell=1)$ calculated from the Lorenz-Mie scattering coefficient $a_1$ and $b_1$ are, respectively,
\begin{align}
\mathbf{p}(a_1) = \varepsilon_0\frac{6\pi\imath a_1}{k^3}\mathbf{E}_0=\mathbf{P} + \imath k\mathbf{T}_{\rm e} + \boldsymbol{\Pi}_{\rm e},\label{Pa1}\\
\mathbf{m}(b_1) = \frac{6\pi\imath b_1}{k^3}\mathbf{H}_0=\mathbf{M} +  k\mathbf{T}_{\rm m} + \boldsymbol{\Pi}_{\rm m},\label{Pb1}
\end{align}
where $(\mathbf{E}_0,\mathbf{H_0})$ is the incident electromagnetic field, with $H_0=E_0/\eta$, and $\boldsymbol{\Pi}_{\rm e,m}$ represents higher-order correcting terms with respect to the size parameter of the sphere~\cite{Miroshnichenko_OptExp23_2015,Miroshnichenko_ACSNano6_2012}.
In the Lorenz-Mie theory, one can only calculate analytically the renormalized scattered powers $W_{\mathbf{p}(a_1)}\propto|\mathbf{p}(a_1)|^2$ and $W_{\mathbf{m}(b_1)}\propto|\mathbf{m}(b_1)|^2$, where the toroidal dipole contributions remain hidden in the scattering quantities~\cite{Alaee_OptCommun2018}.

In the next section, we address the calculation of electric, magnetic, and toroidal dipole excitations in light scattering by core-shell spheres without relying on the far-field dipole approximations provided in Eqs.~(\ref{Erad})--(\ref{Wtm}).
Our goal is to derive analytic expressions for the partial scattering coefficients associated with electric and magnetic toroidal dipole excitations for spheres with size smaller than or comparable to the wavelength, enabling the separation of coefficients $a_1$ and $b_1$ into two primary contributions as described in Eqs.~(\ref{a_E-ap})--(\ref{a_Tm-ap}), and Eqs.~(\ref{Pa1}) and (\ref{Pb1}).

\subsection{Lorenz-Mie theory for core-shell spheres}
\label{Lorenz-Mie}

\begin{figure}[htbp]
\includegraphics[width=\columnwidth]{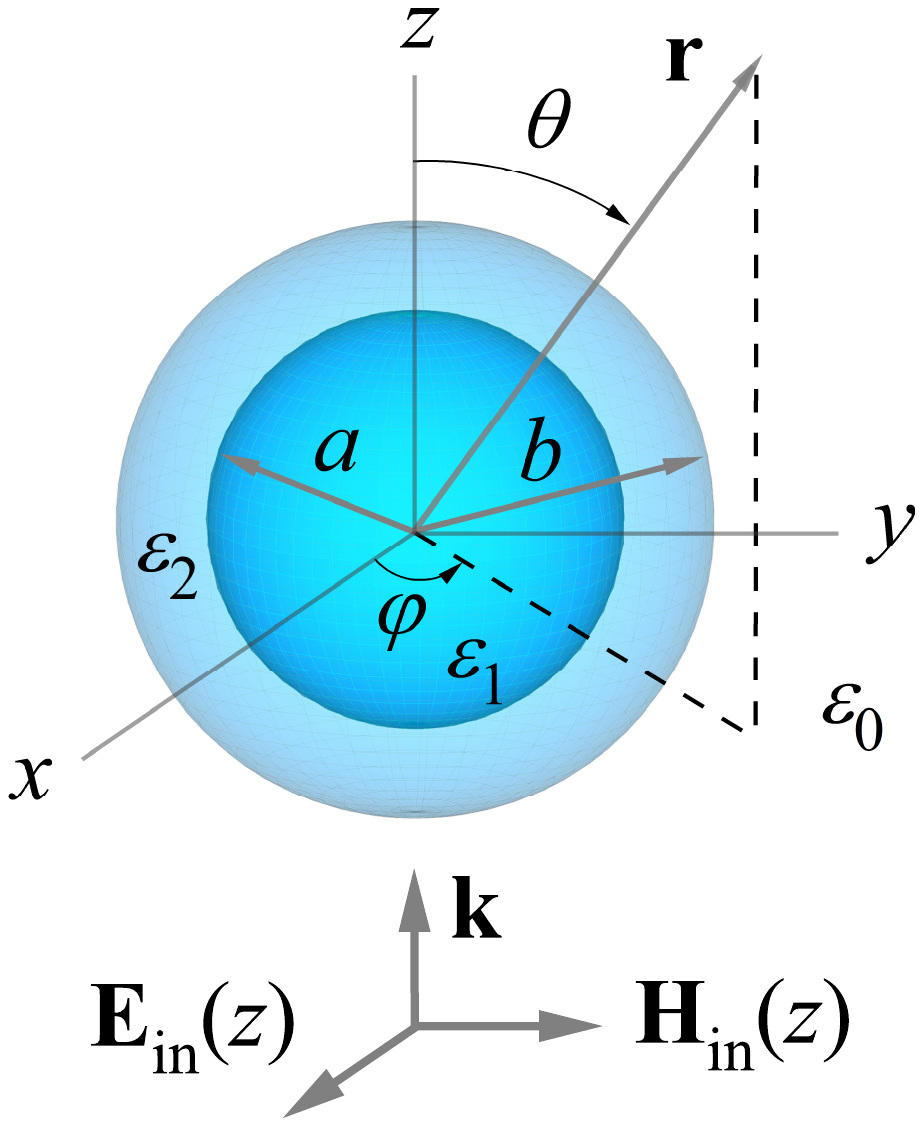}
\caption{A center-symmetric core-shell sphere, with inner radius $a$ and outer radius $b$, irradiated by a linear-polarized electromagnetic plane wave.
The electric permittivities of the core, spherical shell, and surrounding medium are $\varepsilon_1$, $\varepsilon_2$, and $\varepsilon_0$, respectively. }\label{fig2}
\end{figure}

Here, we confine our discussion to the Aden-Kerker extension of the Lorenz-Mie theory~\cite{Bohren_Book_1983,Aden_JAppPhys22_1951}.
We focus on the scattering of electromagnetic plane waves by a core-shell sphere composed of homogeneous, linear, and isotropic materials.
Our aim is to distinguish between electric and magnetic dipole excitations from the toroidal dipole excitations within the framework of the Lorenz-Mie theory.
For this purpose, we examine a nonmagnetic coated sphere of inner radius $a$ and outer radius $b>a$, featuring electric permittivity $\varepsilon_1$ for the core and  $\varepsilon_2$ for the shell, embedded in a lossless medium \color{black} with permittivity \color{black} $\varepsilon_0$, as illustrated in Fig.~\ref{fig2}.

The well-known Lorenz-Mie coefficients $a_{\ell}$ and $b_{\ell}$ can be calculated from Eqs.~(\ref{a_E}) and (\ref{a_M}) provided the source current density $\mathbf{J}(\mathbf{r})$ within the sphere ($|\mathbf{r}|\leq b$).
To calculate $\mathbf{J}(\mathbf{r})$ from Eq.~(\ref{J}), we have to know the electric field distribution inside the sphere.
Without loss of generality, by considering an incident plane wave polarized along the $x$-axis, given by $\mathbf{E}_{\rm in}(z)=E_0\hat{\mathbf{x}}e^{\imath kz}$, the electric field within the core-shell sphere~\cite{Arruda_JOpt14_2012} is
\begin{align}
\mathbf{E}_{1}(\mathbf{r})&=\sum_{\ell=1}^{\infty}\frac{E_{\ell}}{n_1kr}\bigg\{-\imath\cos\varphi\sin\theta d_{\ell} \frac{\psi_{\ell}(n_1kr)}{n_1kr}\ell(\ell+1)\pi_{\ell1}\hat{\bf r}\nonumber\\
&+\cos\varphi\left[c_{\ell}\pi_{\ell1}\psi_{\ell}(n_1kr)-\imath d_{\ell}\tau_{\ell1}\psi_{\ell}'(n_1kr)\right]\hat{\boldsymbol{\theta}}\nonumber\\
&+\sin\varphi\left[\imath d_{\ell}\pi_{\ell1}\psi_{\ell}'(n_1kr)-c_{\ell}\tau_{\ell1}\psi_{\ell}(n_1kr)\right]\hat{\boldsymbol{\varphi}}\bigg\},\label{E1}
\end{align}
for the core ($0<r\leq a$), and
\begin{align}
\mathbf{E}_{2}(\mathbf{r})&=\sum_{\ell=1}^{\infty}\frac{E_{\ell}}{n_2kr}\bigg\{-\imath\cos\varphi\sin\theta g_{\ell} \frac{\psi_{\ell}(n_2kr)}{n_2kr}\ell(\ell+1)\pi_{\ell1}\hat{\bf r}\nonumber\\
&+\imath\cos\varphi\sin\theta w_{\ell} \frac{\chi_{\ell}(n_2kr)}{n_2kr}\ell(\ell+1)\pi_{\ell1}\hat{\bf r}\nonumber\\
&+\cos\varphi\left[f_{\ell}\pi_{\ell1}\psi_{\ell}(n_2kr)-\imath g_{\ell}\tau_{\ell1}\psi_{\ell}'(n_2kr)\right]\hat{\boldsymbol{\theta}}\nonumber\\
&-\cos\varphi\left[v_{\ell}\pi_{\ell1}\chi_{\ell}(n_2kr)-\imath w_{\ell}\tau_{\ell1}\chi_{\ell}'(n_2kr)\right]\hat{\boldsymbol{\theta}}\nonumber\\
&+\sin\varphi\left[\imath g_{\ell}\pi_{\ell1}\psi_{\ell}'(n_2kr)-f_{\ell}\tau_{\ell1}\psi_{\ell}(n_2kr)\right]\hat{\boldsymbol{\varphi}}\nonumber\\
&-\sin\varphi\left[\imath w_{\ell}\pi_{\ell1}\chi_{\ell}'(n_2kr)-v_{\ell}\tau_{\ell1}\chi_{\ell}(n_2kr)\right]\hat{\boldsymbol{\varphi}}
\bigg\},\label{E2}
\end{align}
for the spherical shell ($a< r\leq b$), where $E_{\ell} = E_0\imath^{\ell} (2\ell+1)/[\ell(\ell+1)]$ and $n_q=\sqrt{\varepsilon_q/\varepsilon_0}$ is the relative refractive index of the medium $q=\{1,2\}$ according to Fig.~\ref{fig2}.
The functions $\psi_{\ell}(z)=z j_{\ell}(z)$, $\chi_{\ell}(z)=-z y_{\ell}(z)$, and $\xi_{\ell}(z)=\psi_{\ell}(z)-\imath\chi_{\ell}(z)$ are the Riccati-Bessel, Riccati-Neumann, and Riccati-Hankel functions, respectively, with $j_{\ell}$ and $y_{\ell}$ being the spherical Bessel and Neumann functions~\cite{Bohren_Book_1983}.

The internal Lorenz-Mie coefficients are obtained from boundary conditions and read as~\cite{Bohren_Book_1983,Arruda_JOpt14_2012}:
\begin{align}
        c_{\ell} &=\frac{n_1f_{\ell}}{n_2\psi_{\ell}(n_1ka)}\left[\psi_{\ell}(n_2ka)-B_{\ell}\chi_{\ell}(n_2ka)\right],\label{cn}\\
        d_{\ell} &=\frac{n_1g_{\ell}}{n_2\psi_{\ell}'(n_1ka)}\left[\psi_{\ell}'(n_2ka)-A_{\ell}\chi_{\ell}'(n_2ka)\right],\label{dn}\\
    f_{\ell}&=\frac{\imath
    n_2\left[\psi_{\ell}(n_2kb)-B_{\ell}\chi_{\ell}(n_2kb)\right]^{-1}}{\xi_{\ell}'(kb)-n_2\xi_{\ell}(kb)\mathcal{B}_{\ell}(n_2kb)},\label{fn}\\
    g_{\ell}&=\frac{\imath
    n_2\left[\psi_{\ell}(n_2kb)-A_{\ell}\chi_{\ell}(n_2kb)\right]^{-1}}{n_2\xi_{\ell}'(kb)-\xi_{\ell}(kb)\mathcal{A}_{\ell}(n_2kb)},\label{gn} \\
        v_{\ell} &=B_{\ell}f_{\ell} ,\\
        w_{\ell} &=A_{\ell}g_{\ell} ,\label{wn}
    \end{align}
\color{black}
where the auxiliary functions $\mathcal{A}_{\ell}(n_2kb)$, $\mathcal{B}_{\ell}(n_2kb)$, $A_{\ell}$, and $B_{\ell}$ are provided in Eqs.(\ref{A3})--(\ref{A6}), respectively.
\color{black}

The solution for a homogeneous sphere of radius $b$ can be readily obtained by setting $\varepsilon_1=\varepsilon_2$, i.e., $A_{\ell}=0=B_{\ell}$.
By plugging Eqs.~(\ref{E1}) and (\ref{E2}) into Eq.~(\ref{J}), and using relations Eqs.~(\ref{a_E2}) and (\ref{a_M2}) for $m=1$, one can determine the scattering Lorenz-Mie coefficients $a_{\ell}$ and $b_{\ell}$ after some algebra. For details, see Appendix~\ref{an-and-bn}.

Substituting Eqs.~(\ref{E1}) and (\ref{E2}) into Eq.~(\ref{J}), and using Eqs.~(\ref{a_E-ap})--(\ref{a_Tm-ap}) for $m=1$ and $\ell=1$, we finally obtain after some algebra:
\begin{widetext}
\begin{align}
a_{1}^{\rm c}&=-\frac{2\imath}{3}\left(1-\frac{1}{n_1^2}\right)d_{1}\int_0^{ka}{\rm d}x\left[\psi_{1}(n_1x)+n_1x{\psi_{1}'(n_1x)}\right]\nonumber\\
&-\frac{2\imath}{3}\left(1-\frac{1}{n_2^2}\right)g_{1}\int_{ka}^{kb}{\rm d}x\left\{\psi_{1}(n_2x)-A_{1}\chi_{1}(n_2x)+n_2x\left[\psi_{1}'(n_2x)-A_{1}\chi_{1}'(n_2x)\right]\right\},\label{ae-int1}\\
a_{1}^{\rm t}&=\frac{\imath}{15}\left(1-\frac{1}{n_1^2}\right){d_{1}}\int_0^{ka}{\rm d}x\left[x^2\psi_{1}(n_1x)+2x^2n_1x\psi_1'(n_1x)\right]\nonumber\\
&+\frac{\imath}{15}\left(1-\frac{1}{n_2^2}\right){g_{1}}\int_{ka}^{kb}{\rm d}x\left\{x^2\left[\psi_{1}(n_2x)-A_1\chi_1(n_2x)\right]+2x^2n_2x\left[\psi_1'(n_2x)-A_1\chi_1'(n_2x)\right]\right\},\label{at-int1}\\
b_{1}^{\rm c}&=-\frac{\imath}{3}\left(1-\frac{1}{n_1^2}\right)n_1c_{1}\int_0^{ka}{\rm d}xx^2\psi_{1}(n_1x)-\frac{\imath}{3}\left(1-\frac{1}{n_2^2}\right)n_2f_{1}\int_{ka}^{kb}{\rm d}xx^2{\left[\psi_{1}(n_2x)-B_{1}\chi_{1}(n_2x)\right]},\label{bm-int2}\\
b_{1}^{\rm t}&=\frac{\imath}{30}\left(1-\frac{1}{n_1^2}\right)n_1c_{1}\int_0^{ka}{\rm d}xx^4\psi_{1}(n_1x)+\frac{\imath}{30}\left(1-\frac{1}{n_2^2}\right)n_2f_{1}\int_{ka}^{kb}{\rm d}xx^4{\left[\psi_{1}(n_2x)-B_{1}\chi_{1}(n_2x)\right]},\label{bt-int2}
\end{align}
\end{widetext}
where $a_1^{\rm c}$ is associated with the Cartesian electric dipole contribution, $b_1^{\rm c}$ with the Cartesian magnetic dipole contribution, $a_1^{t}$ with the electric toroidal dipole contribution, and $b_1^{t}$ with the magnetic toroidal dipole contribution.
Here, we have used Eqs.~(\ref{pi1})--(\ref{pi3}) to solve the angular integrals and eliminate the double summation, leaving only radial integrals to be calculated.
The standard Lorenz-Mie scattering coefficients for the small-particle approximation ($\ell=1$ and $kb\ll1$) are the sum of these partial contributions:
\begin{align}
a_{1}&\approx a_{1}^{\rm c}+a_{1}^{\rm t},\label{a1-approx-equal}\\
b_{1}&\approx b_{1}^{\rm c}+b_{1}^{\rm t}.\label{b1-approx-equal}
\end{align}

The solution of the radial integrals for the electric dipole coefficient in Eq.~(\ref{ae-int1}) is  obtained straightforwardly by noting that $\varrho_{1}(x)+x\varrho_{1}'(x)=[x\varrho_{1}(x)]'$, where $\varrho_{1}$ can be either $\psi_1$ or $\chi_1$.
Regarding the electric toroidal coefficient provided in Eq.~(\ref{at-int1}), we apply integration by parts to obtain
$\int {\rm d}x\{x^3\varrho_{1}'(x)+x^2[x\varrho_{1}(x)]'\}=2x^3\varrho_{1}(x)-5\int{\rm d}x x^2\varrho_{1}(x)$.
To solve the remaining integrals, we recall that $z_{\ell}(x)=\sqrt{\pi/2x}Z_{\ell+1/2}(x)$ and $\int x^{\ell}Z_{\ell-1}(x)dx=x^{\ell}Z_{\ell}(x)+C$, where $z_{\ell}$ is any spherical Bessel function ($j_{\ell}$ or $y_{\ell}$) and $Z_{\ell}$ is any cylindrical Bessel function ($J_{\ell}$ or $Y_{\ell}$)~\cite{Abramovitz_1964}.
By manipulating these expressions, we obtain the analytic solution
\begin{align}
\int_{l_1}^{l_2} x^{\ell+2} z_{\ell}(x)dx &= x^{\ell+2}z_{\ell+1}(x)\bigg|_{x=l_1}^{x=l_2},\label{int1-jn}
\end{align}
where $z_{\ell}$ is any spherical Bessel function, $j_{\ell}$ or $y_{\ell}$, defined on the closed interval $[l_1,l_2]$.

Applying the fundamental theorem of calculus to Eq.~(\ref{ae-int1}), and utilizing Eq.~(\ref{int1-jn}) in Eq.~(\ref{at-int1}) after integrating by parts, we finally obtain the analytic expressions for the electric dipole and toroidal dipole coefficients:

\begin{widetext}
\begin{align}
a_{1}^{\rm c}
&=\frac{2\imath ka}{3}\left(\frac{1}{n_1^2}-\frac{1}{n_2^2}\right)g_{1}\left[\psi_{1}(n_2ka)-A_{1}\chi_{1}(n_2ka)\right]-\frac{2\imath kb}{3}\left({1}-\frac{1}{n_2^2}\right)g_{1}\left[\psi_{1}(n_2kb)-A_{1}\chi_{1}(n_2kb)\right],\label{ae-analytic}\\
a_{1}^{\rm t}&=\frac{\imath(ka)^3}{15}\left(1-\frac{1}{n_1^2}\right)g_{1}\left[1-\frac{\psi_{3}(n_1ka)}{\psi_1(n_1ka)}\right]\left[\psi_{1}(n_2ka)-A_{1}\chi_{1}(n_2ka)\right]\nonumber\\
&-\dfrac{\imath(ka)^3}{15}\left(1-\frac{1}{n_2^2}\right)g_{1}\bigg\{\psi_1(n_2ka)-\psi_{3}(n_2ka)-A_1\left[\chi_1(n_2ka)-\chi_3(n_2ka)\right]\bigg\}\nonumber\\
&+\dfrac{\imath(kb)^3}{15}\left(1-\frac{1}{n_2^2}\right) g_{1}\bigg\{\psi_1(n_2kb)-\psi_{3}(n_2kb)-A_1\left[\chi_1(n_2kb)-\chi_3(n_2kb)\right]\bigg\},\label{at-analytic}
\end{align}
\end{widetext}
where we have used the boundary condition equation between the internal coefficients $d_{\ell}$ and $g_{\ell}$ to simplify the final expression~\cite{Bohren_Book_1983}: $\psi_{\ell}(n_1ka) d_{\ell}= g_{\ell}[\psi_{\ell}(n_2ka)-A_{\ell}\chi_{\ell}(n_2ka)]$.

In addition, using integration by parts and Eq.~(\ref{int1-jn}), we can also obtain
\begin{align}
\int_{l_1}^{l_2} x^{\ell+4} z_{\ell}(x)dx &= x^{\ell+3}\left[xz_{\ell+1}(x)-{2}z_{\ell+2}(x)\right]\bigg|_{x=l_1}^{x=l_2}.\label{int2-jn}
\end{align}
Applying the result of Eq.~(\ref{int1-jn}) to Eq.~(\ref{bm-int2}), and using Eq.~(\ref{int2-jn}) in Eq.~(\ref{bt-int2}), we obtain the magnetic dipole and magnetic toroidal dipole coefficients:
\begin{widetext}
\begin{align}
b_{1}^{\rm c}
&=-\frac{\imath (ka)^2}{3}\left({1}-\frac{1}{n_1^2}\right)\frac{n_1}{n_2}f_1\frac{\psi_2(n_1ka)}{\psi_1(n_1ka)}\left[\psi_1(n_2ka)-B_1\chi_1(n_2ka)\right]\nonumber\\
&+\frac{\imath (ka)^2}{3}\left(1-\frac{1}{n_2^2}\right)f_1\left[\psi_2(n_2ka)-B_1\chi_2(n_2ka)\right]-\frac{\imath (kb)^2}{3}\left(1-\frac{1}{n_2^2}\right)f_1\left[\psi_2(n_2kb)-B_1\chi_2(n_2kb)\right],\label{bm-analytic}\\
b_{1}^{\rm t}&=\frac{\imath (ka)^4}{30}\left({1}-\frac{1}{n_1^2}\right)\frac{n_1}{n_2}f_1\frac{\left[\psi_2(n_1ka)-2j_3(n_1ka)\right]}{\psi_1(n_1ka)}\left[\psi_1(n_2ka)-B_1\chi_1(n_2ka)\right]\nonumber\\
&-\frac{\imath (ka)^4}{30}\left({1}-\frac{1}{n_2^2}\right)f_1\bigg\{\psi_2(n_2ka)-2j_3(n_2ka)-B_1\left[\chi_2(n_2ka)+2y_3(n_2ka)\right]\bigg\}\nonumber\\
&+\frac{\imath (kb)^4}{30}\left({1}-\frac{1}{n_2^2}\right)f_1\bigg\{\psi_2(n_2kb)-2j_3(n_2kb)-B_1\left[\chi_2(n_2kb)+2y_3(n_2kb)\right]\bigg\},\label{bt-analytic}
\end{align}
\end{widetext}
where we have used the boundary condition equation $\psi_{\ell}'(n_1ka) c_{\ell}=f_{\ell}[\psi_{\ell}'(n_2ka)-B_{\ell}\chi_{\ell}'(n_2ka)]$~\cite{Bohren_Book_1983}.

Equations~(\ref{ae-analytic})--(\ref{bt-analytic}) represent the main analytic results of this paper and can be applied to small particles (compared to the wavelength) beyond the Rayleigh limit, as we do not impose $n_1kb,\ n_2kb\ll 1$.
Indeed, these equations allow for the direct calculation of toroidal dipole contributions in light scattering by core-shell spheres for $kb\approx 1$, without relaying on numerical integration of the local charge-current density.
In particular, the classical limiting case of a homogeneous sphere of radius $R$ and relative refractive index $n$ (in relation to the surrounding medium) can be readily obtained by setting $n_1=n_2=n$ and $b=R$ in Eqs.~(\ref{ae-analytic})--(\ref{bt-analytic}):
\begin{align}
a_{1}^{\rm c}
&=-\frac{2\imath kR}{3}\left({1}-\frac{1}{n^2}\right)d_{1}\psi_{1}(nkR),\\
a_{1}^{\rm t}&=\dfrac{\imath(kR)^3}{15}\left(1-\frac{1}{n^2}\right) d_{1}\left[\psi_1(nkR)-\psi_{3}(nkR)\right],\\
b_{1}^{\rm c}
&=-\frac{\imath (kR)^2}{3}\left(1-\frac{1}{n^2}\right)c_1\psi_2(nkR),\\
b_{1}^{\rm t}&=\frac{\imath (kR)^4}{30}\left({1}-\frac{1}{n^2}\right)c_1\left[\psi_2(nkR)-2j_3(nkR)\right],
\end{align}
where the internal coefficients for a homogeneous sphere considering $\ell=1$ are~\cite{Arruda_JOSA27_2010}
\begin{align}
c_1&=\frac{\imath n}{\psi_1(nkR)\xi_1'(kR)-n\psi_1'(nkR)\xi_1(kR)},\\
d_1&=\frac{\imath n}{n\psi_1(nkR)\xi_1'(kR)-\psi_1'(nkR)\xi_1(kR)}.
\end{align}

\section{Modulating toroidal dipole excitations in core-shell spheres}
\label{Numerical}

Let us consider the scattering properties of a core-shell nanosphere in two different configurations:
(i) a plasmonic silver nanosphere coated by a gain-assisted dielectric nanoshell,
and (ii) a silver nanoshell coating a gain-assisted dielectric core.
Our aim is to excite both electric and magnetic toroidal dipole resonances within a certain range of parameters using plasmonic core-shell scatterers containing a gain medium.
\color{black}
Experimentally, core-shell nanostructures can be synthesised based on lithographic techniques (e.g., electron or ion beam), mechanical approaches, chemical synthesis (e.g., self-assembly, film deposition, and growth), among other synthesis mechanisms~\cite{Paria_ChemRev112_2012}.
\color{black}

Typically, gain-assisted materials can consist of dielectric media doped with dye molecules or rare-earth ions.
In these materials, there is an inversion in the number of electrons, resulting in a higher population in the excited level compared to the lower level, which leads to a negative imaginary part of the refractive index~\cite{Tsakmakidis_Science339_2013}.
Here, we choose doped AlGaAs for the linear-gain material, with its refractive index at optical frequencies approximated by $n_{\rm AlGaAs}=3.5-\imath\kappa$, where $\kappa$ is a phenomenological gain coefficient (for further details, see Ref.~\cite{Nano} and references therein).
This linear gain approximation for $n_{\rm AlGaAs}$ is valid at or below threshold~\cite{Soukoulis_PhysRevB59_1999}.
For the plasmonic material composing the scatterer we consider silver (Ag), whose permittivity is well described by the Drude model
\begin{align}
\frac{\varepsilon_{\rm Ag}(\omega)}{\varepsilon_0}=\varepsilon_{\rm int}-\frac{\omega_{\rm p}^2}{\omega(\omega+\imath\gamma)},
\end{align}
where $\varepsilon_{\rm int}= 3.7$ is a contribution due to interband transitions, $\omega_{\rm p}= 9.2$~eV
($\approx 2\pi \times 2.2 \times 10^{15}$~Hz) is the plasmon frequency associated with conduction electrons,
and $\gamma = 0.02$~eV is the effective dumping rate due to material losses. These
Drude parameters reproduce experimental data and are valid below the frequency of onset for interband transitions:
$\omega < 0.42\ \omega_{\rm p}$~\cite{Boltasseva_LaserPhotRev4_2010,Christy_PhysRevB6_1972}.

In the full-wave Lorenz-Mie theory, the scattering properties of a spherical particle can be  quantified analytically by the total scattering cross section $\sigma_{\rm sca}$, which is calculated from~\cite{Bohren_Book_1983}
\begin{align}
\sigma_{\rm sca} &= \frac{2\pi}{k^2}\sum_{\ell=1}^{\infty}(2\ell+1)\left(|a_{\ell}|^2+|b_{\ell}|^2\right),
\end{align}
where the analytic expressions for the scattering coefficients $a_{\ell}$ and $b_{\ell}$ are provided in Appendix~\ref{an-and-bn}.
For our purposes, we define the partial contribution to the scattering cross section at the first-order excitation ($\ell=1$) as
\begin{align}
\sigma(\alpha)=\frac{6\pi}{k^2}|\alpha|^2,\label{sigma-partial}
\end{align}
\color{black}
where $\alpha$ can be either $a_1$ (spherical electric dipole, s-ED), $a_1^{\rm c}$ (Cartesian electric dipole, c-ED), $a_1^{\rm t}$ (electric toroidal dipole, ETD), $b_1$ (spherical magnetic dipole, s-MD), $b_1^{\rm c}$ (Cartesian magnetic dipole, c-MD), or $b_1^{\rm t}$ (magnetic toroidal dipole, MTD).
Henceforth, s-ED and s-MD refer to the exact spherical electric and magnetic dipoles provided by the Lorenz-Mie theory; all the other acronyms refer to the Cartesian multipoles.
\color{black}

\subsection{Comparison between the partial scattering coefficients and Cartesian toroidal dipoles}

\begin{figure}[htbp]
\includegraphics[width=.9\columnwidth]{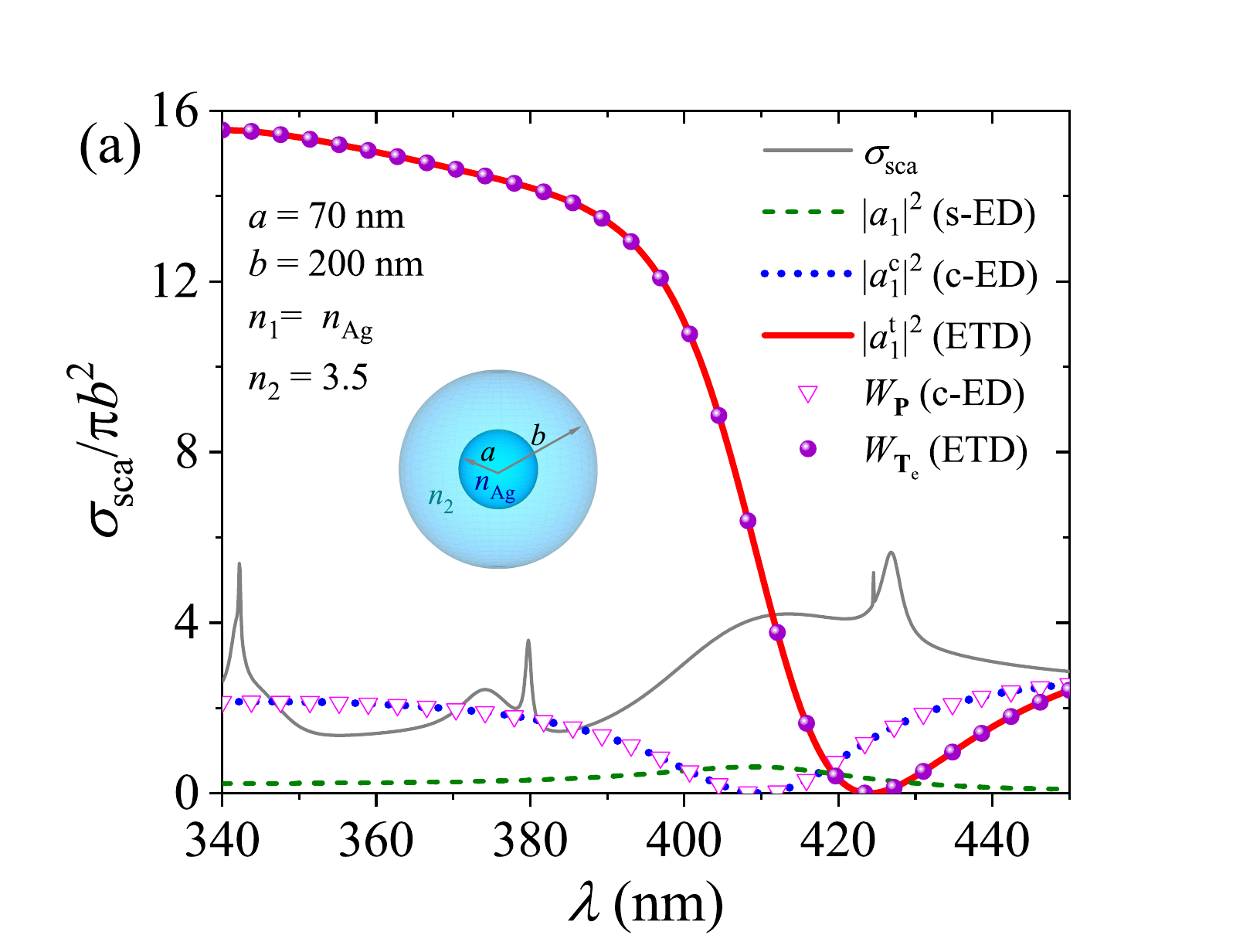}\\
\vspace{-.5cm}
\includegraphics[width=.9\columnwidth]{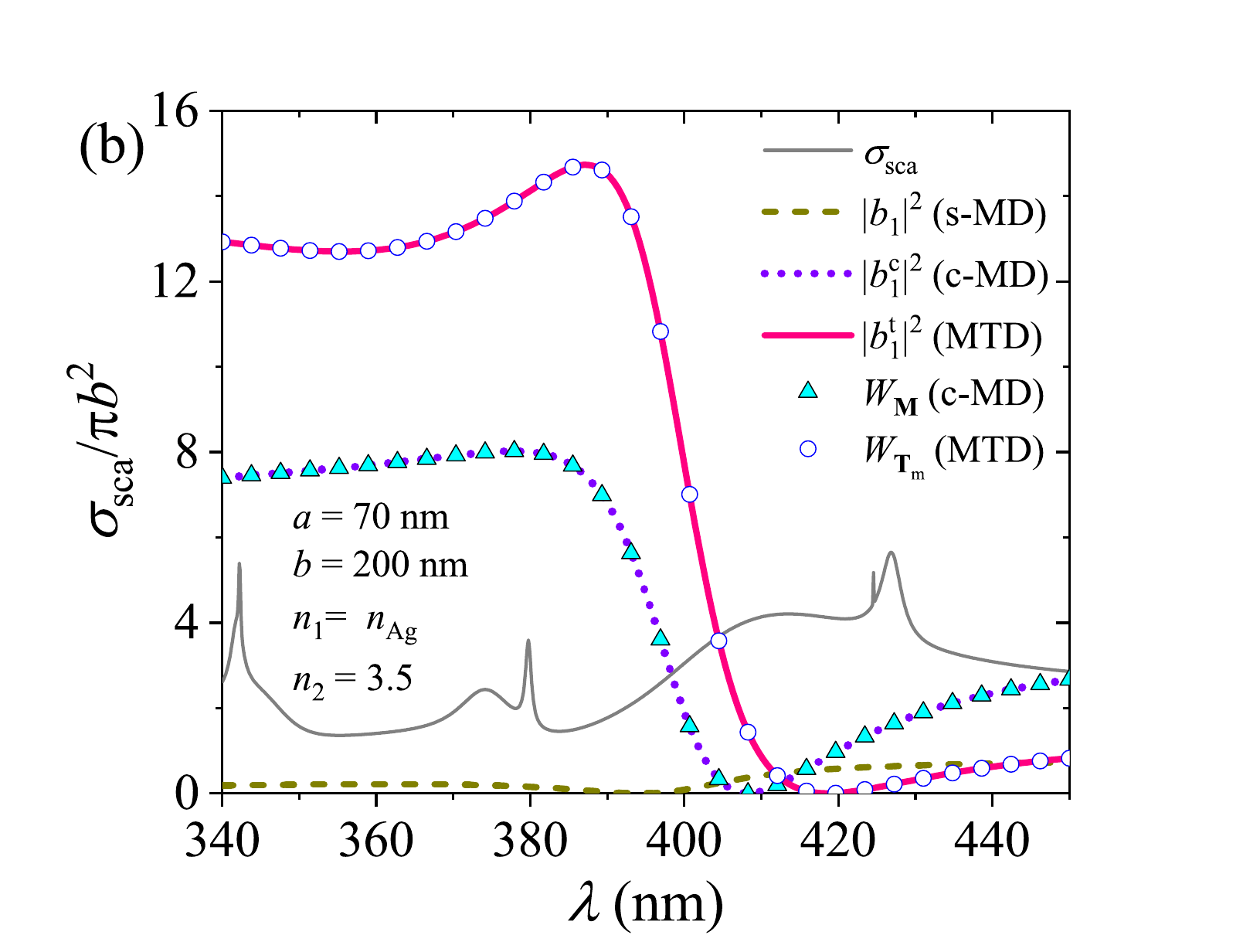}
\caption{Scattering cross section $\sigma_{\rm sca}$ (in units of $\pi b^2$) and first-order excitations associated with an (Ag) core-shell (dielectric) nanosphere of inner radius $a=70$~nm and outer radius $b= 200$~nm, with $n_2=3.5$ the refractive index of the spherical shell.
(a) The electric first-order contributions related to $a_1$ (spherical electric dipole, s-ED), $a_1^{\rm c}$ (Cartesian electric dipole, c-ED), $a_1^{\rm t}$ (electric toroidal dipole, ETD) are compared with $W_{\bf P}$ (c-ED) and $W_{{\bf T}_{\rm e}}$ (ETD).
(b) The magnetic first-order contributions related to $b_1$ (spherical magnetic dipole, s-MD), $b_1^{\rm c}$ (Cartesian magnetic dipole, c-MD), $b_1^{\rm t}$ (magnetic toroidal dipole, MTD) are compared with $W_{\bf M}$ (c-MD) and $W_{{\rm T}_{\rm m}}$ (MTD).
}\label{fig3}
\end{figure}

Before delving into the study of toroidal dipole excitations in core-shell nanoparticles, it is convenient to demonstrate that the closed-form analytic expressions derived for $a_1^{\rm c}$, $a_1^{\rm t}$, $b_1^{\rm c}$, and $b_1^{\rm t}$, as provided in Eqs.~(\ref{ae-analytic}), (\ref{at-analytic}), (\ref{bm-analytic}) and (\ref{bt-analytic}), respectively, upon insertion into Eq.~(\ref{sigma-partial}), produce consistent results when compared to the well-established Eqs.~(\ref{Wp})--(\ref{Wtm}), which are computed through numerical integration of fields and source currents across the scatterer's volume.
With this aim, let us first consider a plasmonic sphere of radius $a=70$~nm and refractive $n_1=n_{\rm Ag}(\omega)$ coated with a dielectric nanoshell of radius $b=200$~nm (i.e., layer thickness $b-a=130$~nm) and refractive index $n_2=3.5$ in free space, as depicted in Fig.~\ref{fig2}.
For incident electromagnetic waves with wavelengths in the range of $340$ to $450$~nm, the size parameter $kb$ of this (Ag) core-shell (dielectric) nanosphere varies within the range $2.7< kb< 3.7$, which is far beyond the small-particle limit ($kb\ll1$).
In this range of parameters, there are higher-order multipole contributions to light scattering that makes the Cartesian dipole approximations given in Eqs.~(\ref{Wp})--(\ref{Wtm}) physically inconsistent as they are valid for $kb\approx 1$ and below, whereas the spherical dipoles are exactly calculated in the Mie regime.
However, we can still use Eq.~(\ref{sigma-partial}) to verify if the two approaches, the numerical integration and the analytic expressions derived in the later section, yield the same results since toroidal dipoles in light scattering can only be excited beyond the electrostatic regime.
In fact, we specifically choose this range of parameters to be able to excite and compare the c-ED, ETD, c-MD, and MTD in both approaches simultaneously, which is only possible beyond the small-particle limit.

Figure~\ref{fig3} shows the scattering cross section $\sigma_{\rm sca}$ normalized to the geometrical cross section, $\sigma_{\rm g}=\pi b^2$, as a function of the incident wavelength $\lambda$.
The partial contributions to electromagnetic scattering associated with the first-order excitations $(\ell=1)$, using Eq.~(\ref{sigma-partial}), are also included in the plots.
In Fig.~\ref{fig3}(a), we compare the numerical and analytical approaches to calculate the c-ED and ETD contributions to light scattering by a core-shell sphere.
In particular, the scattered powers associated with the Cartesian dipoles defined in  Eqs.~(\ref{Wp})--(\ref{Wtm}) are normalized to the incident field intensity, $I_0=(1/2)E_0^2\sqrt{\epsilon_0/\mu_0}$, and to the geometrical cross section, $\sigma_{\rm g}=\pi b^2$.
As can be verified in Fig.~\ref{fig3}(a), the plots for the Cartesian electric dipole $W_{\bf P}$ (bullets) and the Cartesian electric toroidal dipole $W_{{\bf T}_{\rm e}}$ (triangles) provides the same result as the analytic expressions associated with $\sigma(a_1^{\rm c})\propto|a_1^{\rm c}|^2$ (c-ED, dotted line) and $|a_1^{\rm t}|^2$ (ETD, solid line).
Similarly, the same behavior is also verified for the magnetic contributions in Fig.~\ref{fig3}(b).
The plots for the Cartesian magnetic dipole $W_{\bf M}$ (circles) and the Cartesian magnetic toroidal dipole $W_{{\bf T}_{\rm m}}$ (triangles) are in perfect agreement with the plots associated with $\sigma(b_1^{\rm c})\propto|b_1^{\rm c}|^2$ (c-MD, dotted line) and $|b_1^{\rm t}|^2$ (MTD, solid line), respectively.
Therefore, Figs.~\ref{fig3}(a) and \ref{fig3}(b) demonstrate that one can choose between one of these two approaches to calculate the Cartesian toroidal dipole excitations within the framework of the Lorenz-Mie theory.
In particular, as the expressions are taken far beyond the small-particle limit, the spherical (renormalized) electric and magnetic dipoles associated with the Lorenz-Mie coefficients $a_1$ and $b_1$ (dashed lines), respectively, cannot be retrieved by the partial coefficients as in Eqs.~(\ref{a1-approx-equal}) and (\ref{b1-approx-equal}) due to higher-order contributions, \color{black} as explicitly stated in Eqs.~(\ref{Pa1}) and (\ref{Pb1}).
Indeed, for the chosen parameters, the c-ED, c-MD, ETD, and MTD cannot be analysed alone as first-order excitations due to correcting terms that must be taken into account when compared to the exact expressions for the s-ED and s-MD~\cite{Alaee_OptCommun2018}.
\color{black}

\subsection{Electric toroidal dipole in a plasmonic sphere coated with a dielectric shell}
\label{Ag-core}

\begin{figure*}[htbp]
\hspace{-.9cm}
\includegraphics[width=.45\textwidth]{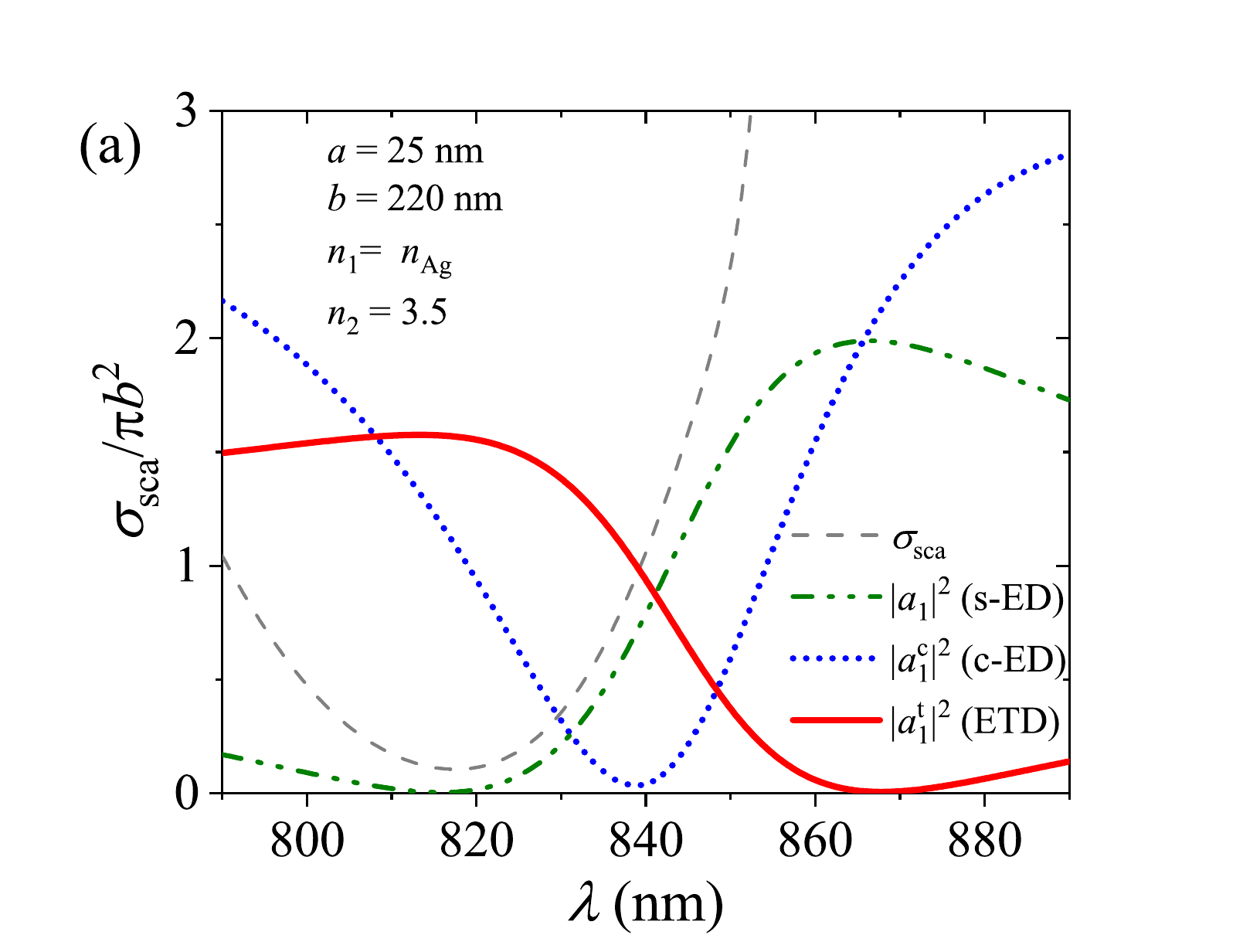}
\includegraphics[width=.45\textwidth]{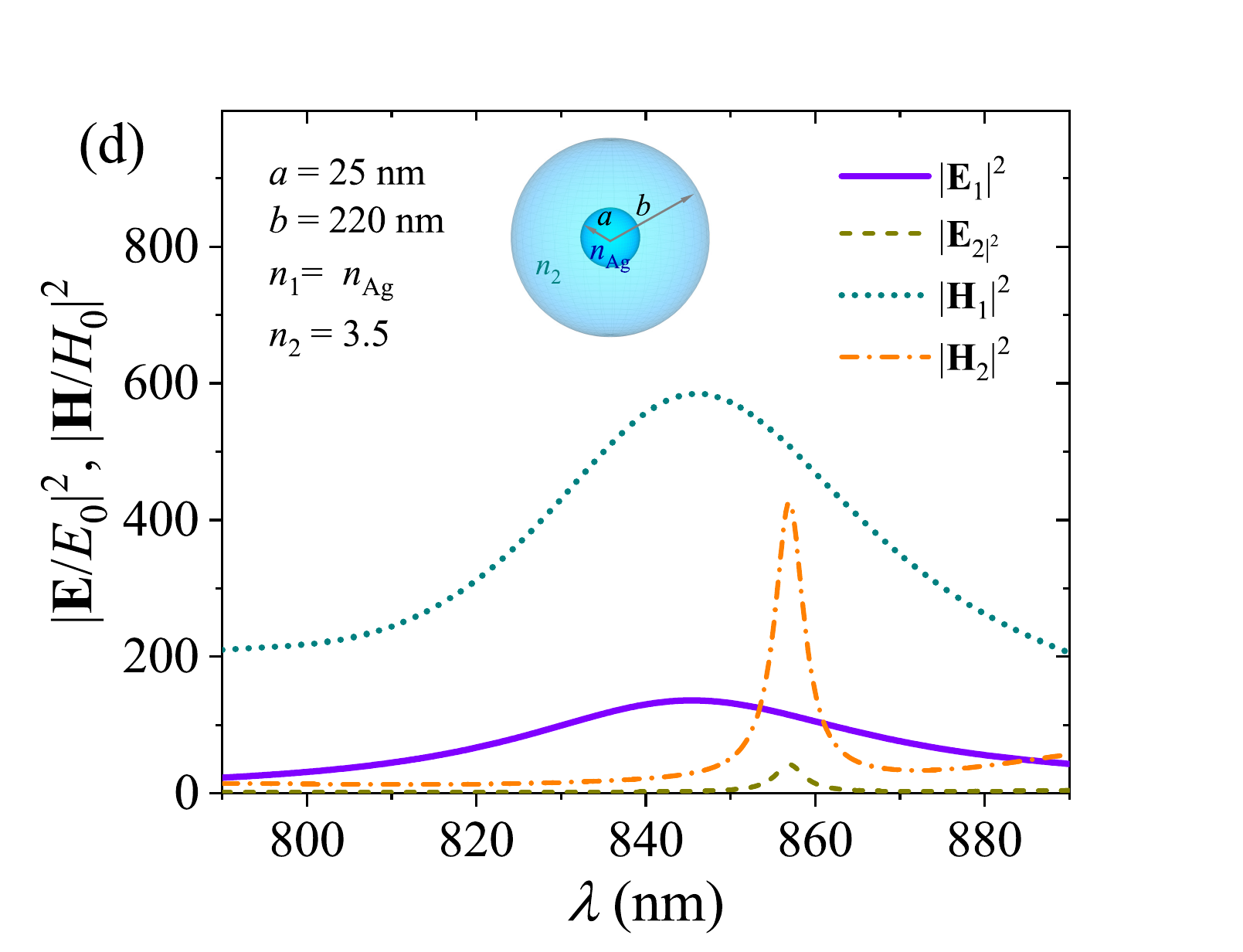}\\
\vspace{-.5cm}
\hspace{-.9cm}
\includegraphics[width=.45\textwidth]{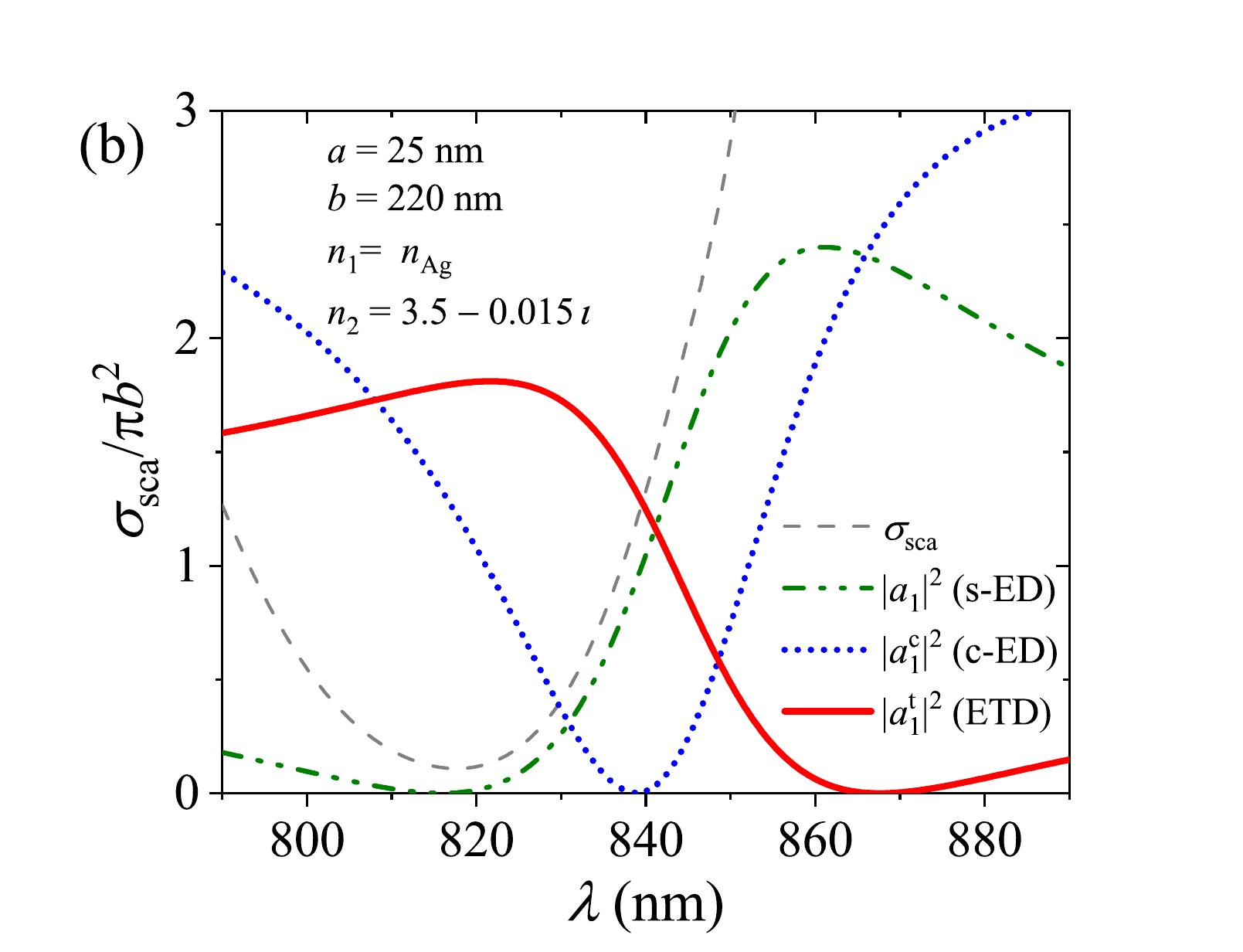}
\includegraphics[width=.45\textwidth]{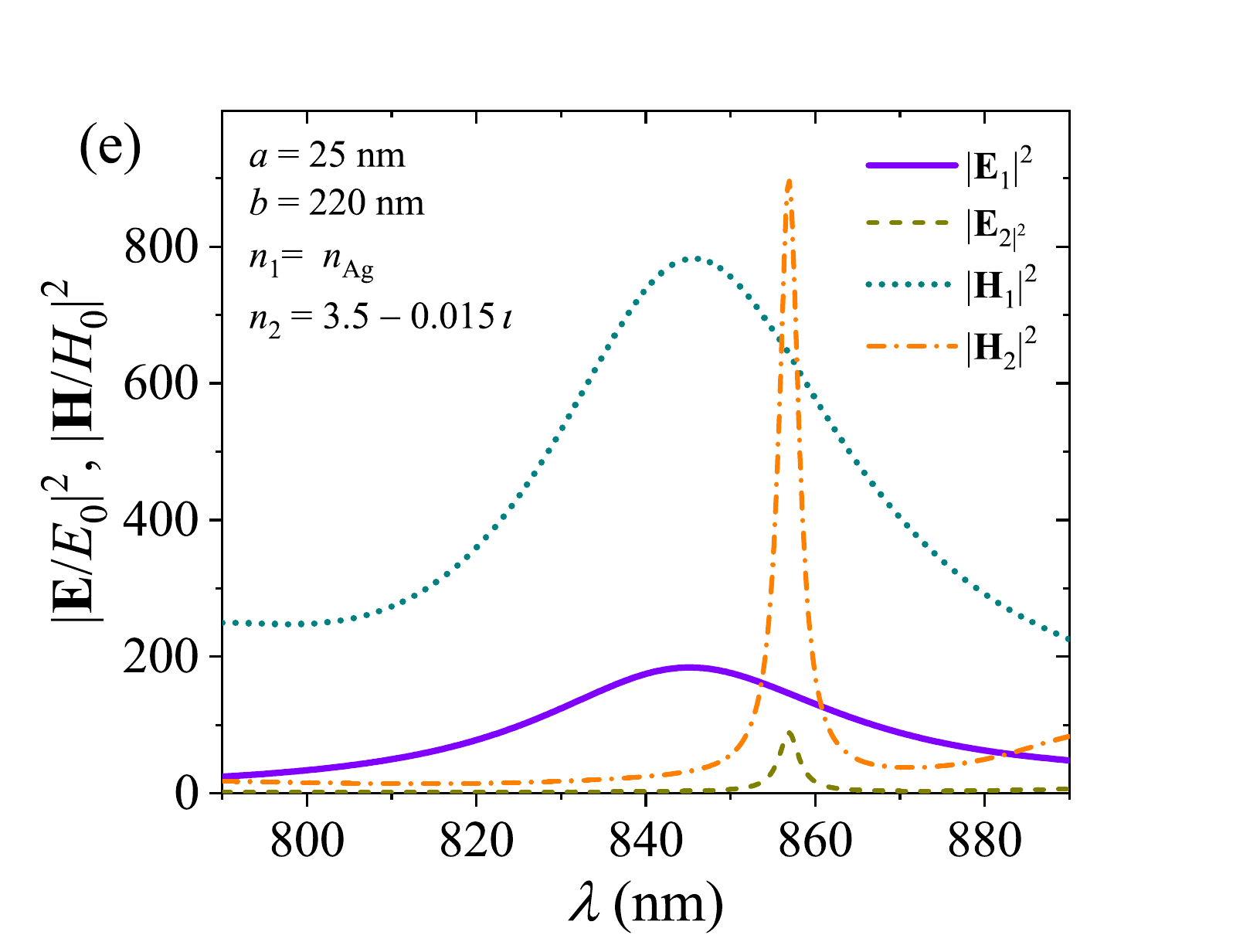}\\
\vspace{-.5cm}
\hspace{-.9cm}
\includegraphics[width=.45\textwidth]{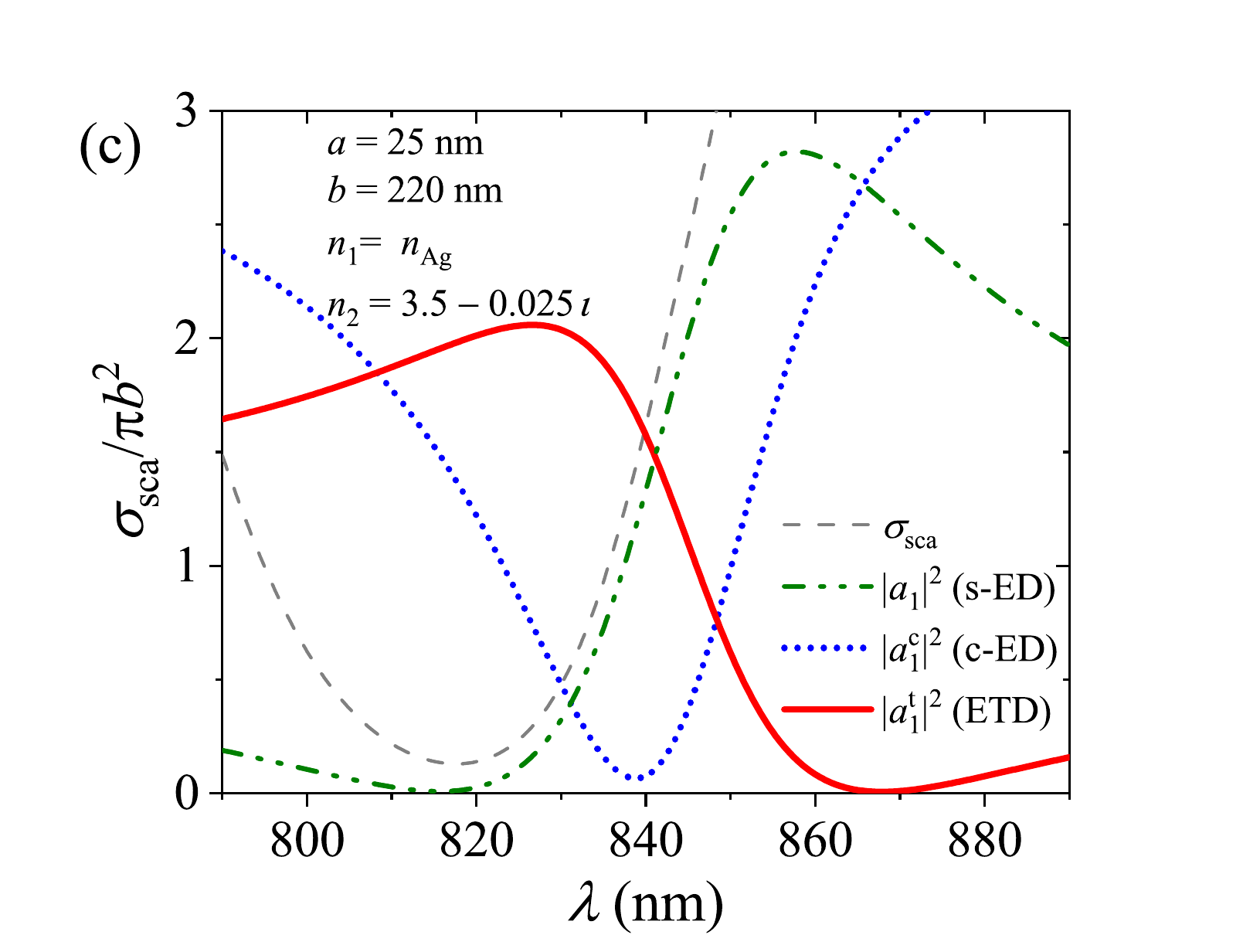}
\includegraphics[width=.45\textwidth]{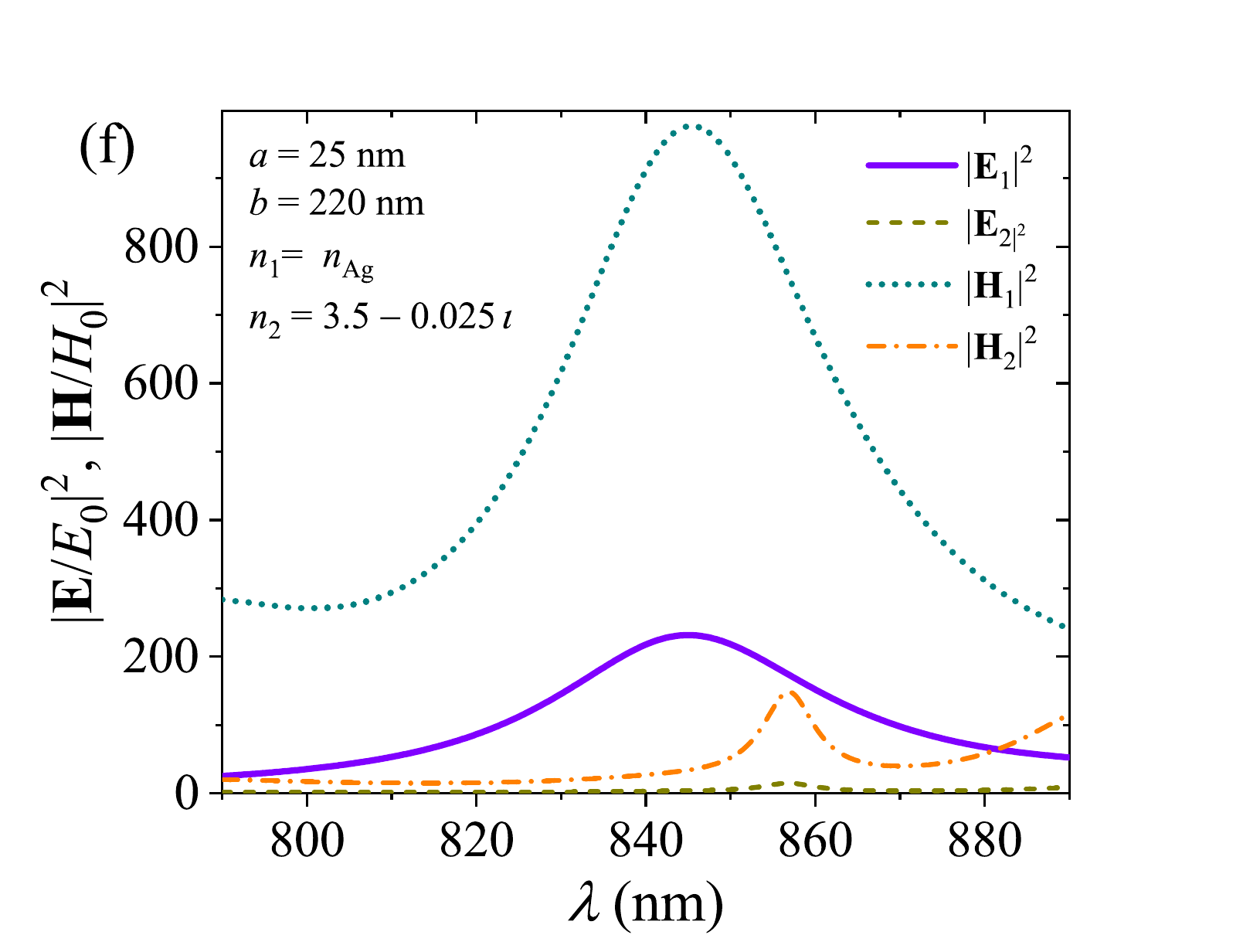}
\caption{
A silver  nanosphere of radius $a=25$~nm coated with an AlGaAs nanoshell $(n_2=3.5-\imath \kappa)$ of radius $b=220$~nm, with $\kappa$ a gain coefficient. The scattering cross-section $\sigma_{\rm sca}$ (dashed line) and the contributions of the spherical electric dipole associated with $a_1$ (s-ED, dashed-dot-dot line), the Cartesian electric dipole with $a_1^{\rm c}$ (c-ED, dotted line), and the electric toroidal dipole with $a_1^{\rm t}$ (ETD, solid line) are plotted as a function of the wavelength for (a) $\kappa=0$, (b) $\kappa=0.015$, and (c) $\kappa=0.025$. The corresponding electric $|\mathbf{E}_q|^2$ and magnetic $|\mathbf{H}_q|^2$ field intensities inside the plasmonic core ($q=1$) and dielectric shell $(q=2)$ are also shown for (d) $\kappa=0$, (e) $\kappa=0.015$, and (f) $\kappa=0.025$.
}\label{fig4}
\end{figure*}

\begin{figure}[htbp]
\includegraphics[width=0.9\columnwidth]{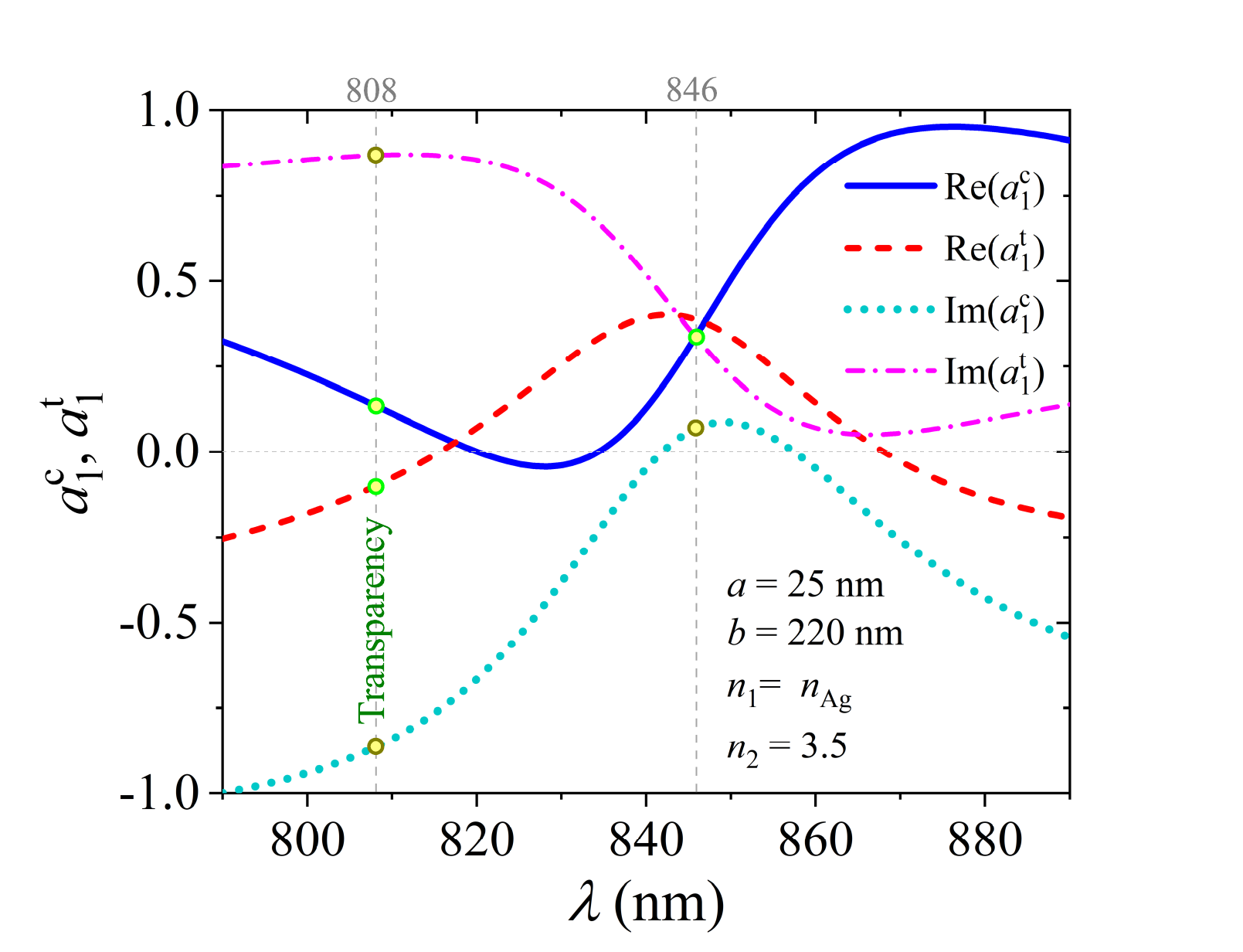}
\caption{First-order scattering coefficients associated with light scattering by an Ag sphere of radius $a=70$~nm coated with a dielectric shell ($n_2=3.5$) of radius $b= 200$~nm.
The plots show the real and imaginary parts of the partial coefficients $a_1^{\rm c}$ and $a_1^{\rm t}$, which are related to the Cartesian electric dipole and electric toroidal dipole, respectively.
}\label{fig5}
\end{figure}

\begin{figure*}[htbp]
\hspace{-.9cm}
\includegraphics[width=.45\textwidth]{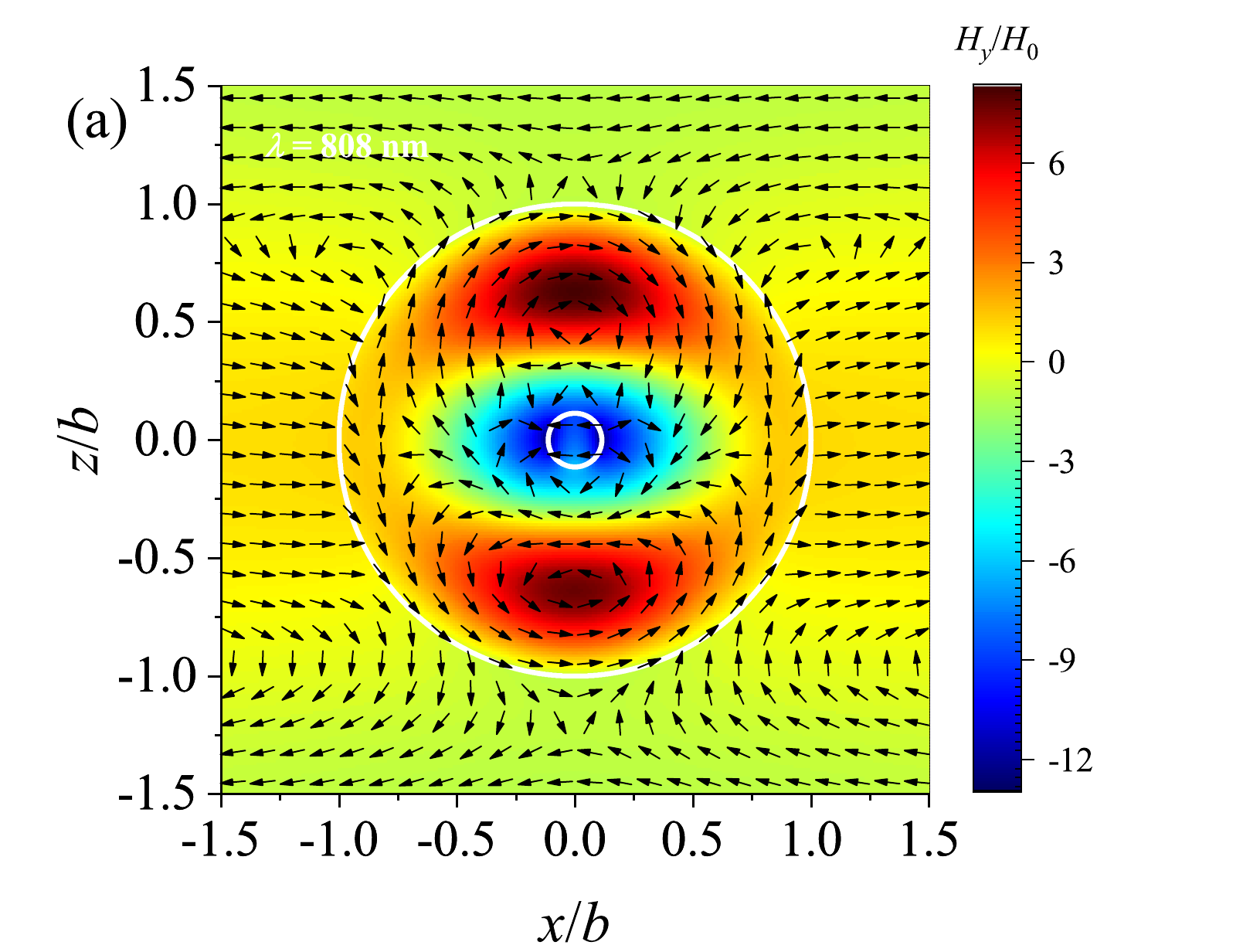}\hspace{-1cm}
\includegraphics[width=.45\textwidth]{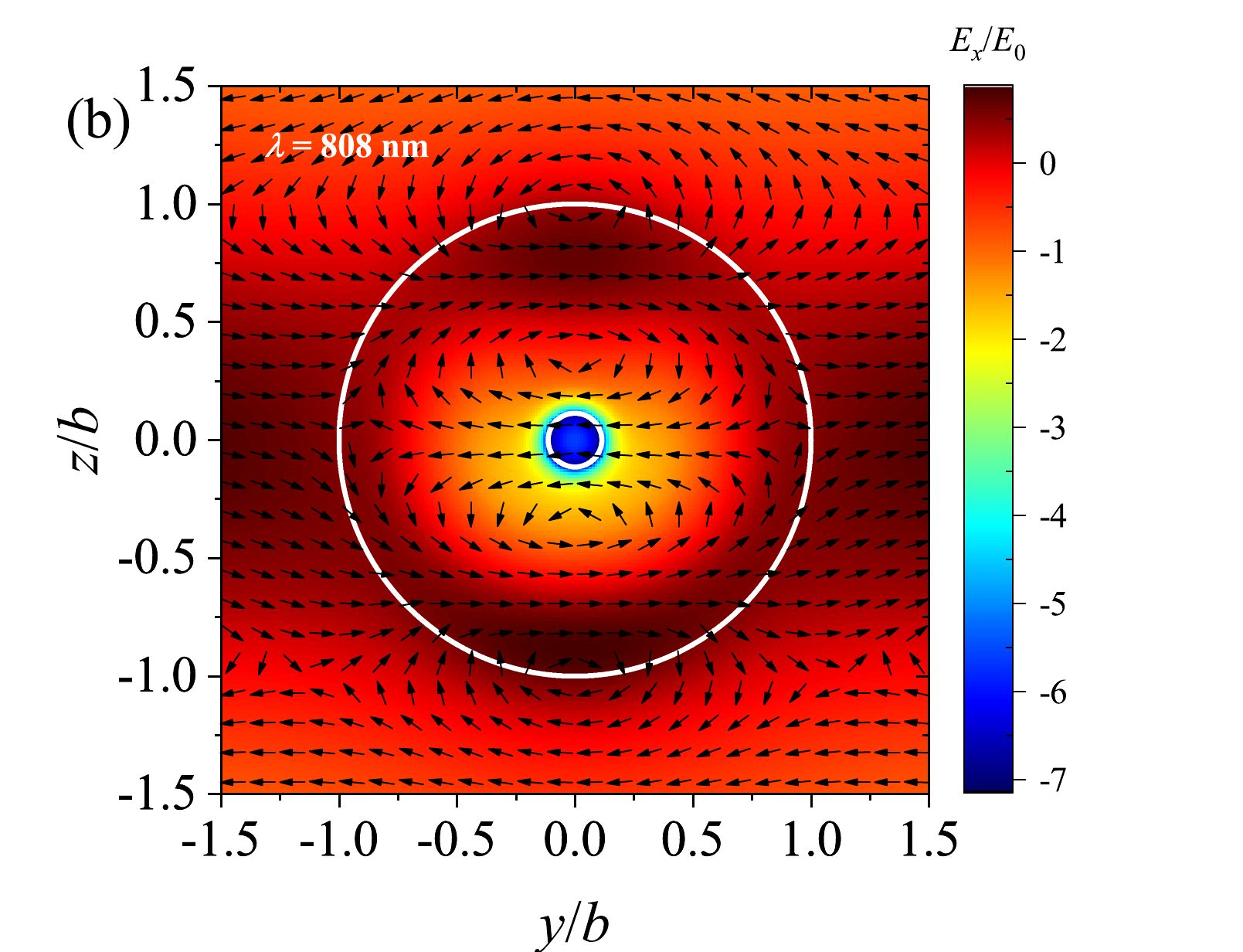}\\
\vspace{-.5cm}
\hspace{-.9cm}
\includegraphics[width=.45\textwidth]{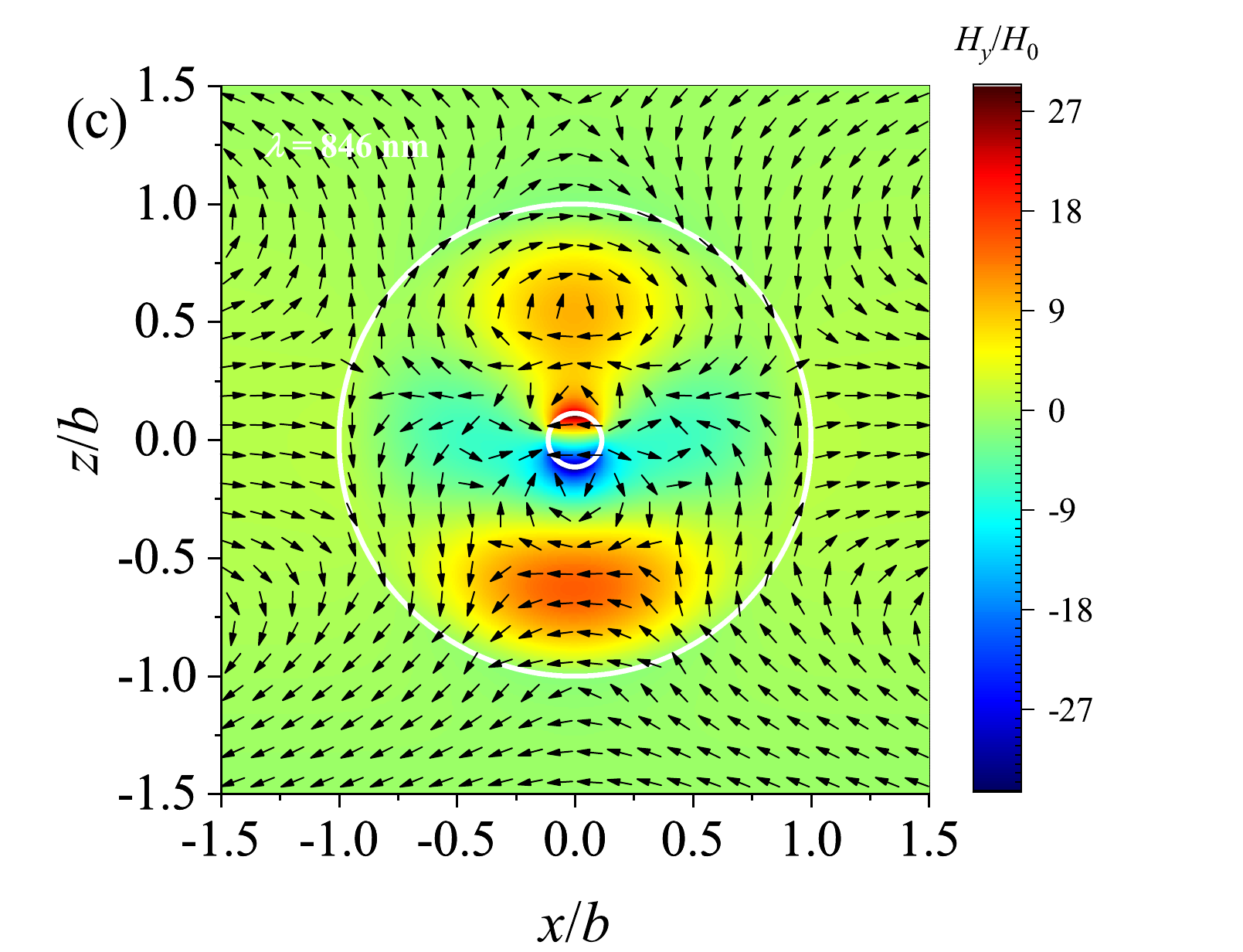}\hspace{-1cm}
\includegraphics[width=.45\textwidth]{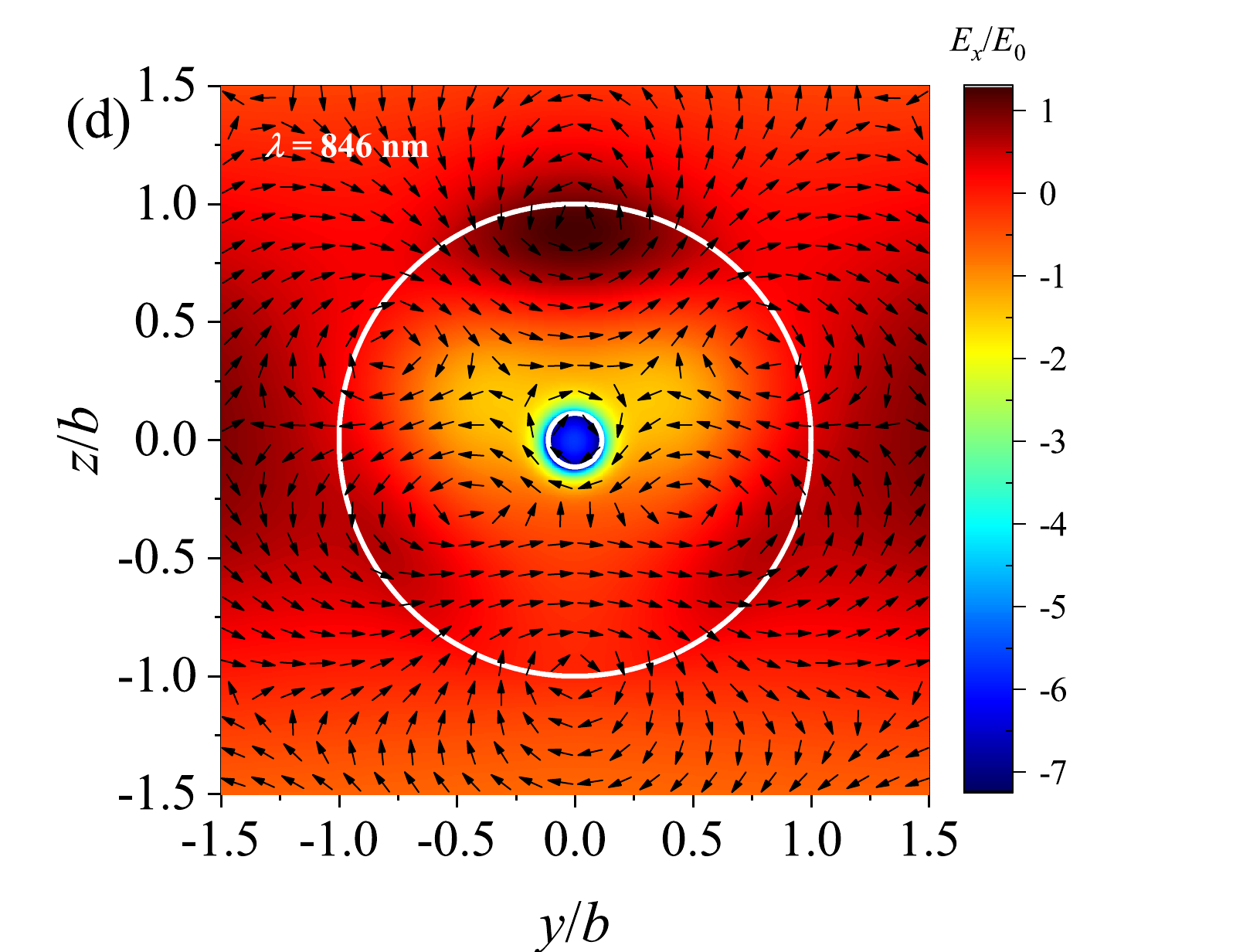}
\caption{Near-field distributions for a silver sphere of radius $a=25$~nm coated with a dielectric shell $(n_2=3.5$) of radius $b=220$~nm.
The components $H_y$ (magnetic field along the $y$-direction, color plots) are calculated from the Lorenz-Mie theory for (a) $\lambda=808$~nm and (c) $\lambda=846$~nm in the $xz$ plane passing through the center of the sphere, where the two concentric white circles delimitates the core-shell sphere.
The wavelength $\lambda=808$~nm corresponds to toroidal dipole-induced transparency in Fig.~\ref{fig4}(a), whereas $\lambda=846$~nm leads to field intensity resonance inside the core in Fig.~\ref{fig4}(d).
The normalized electric vector field in $xz$ plane is represented by black arrows (vector plots).
The components $E_x$ (electric field along the $x$-direction, color plots) are calculated for (b) $\lambda=808$~nm and (d) $\lambda=846$~nm in the $yz$ plane passing through the center of the sphere, where the magnetic vector field in $yz$ plane is represented by black arrows (vector plots).
}\label{fig6}
\end{figure*}

\begin{figure}[htbp]
\includegraphics[width=0.9\columnwidth]{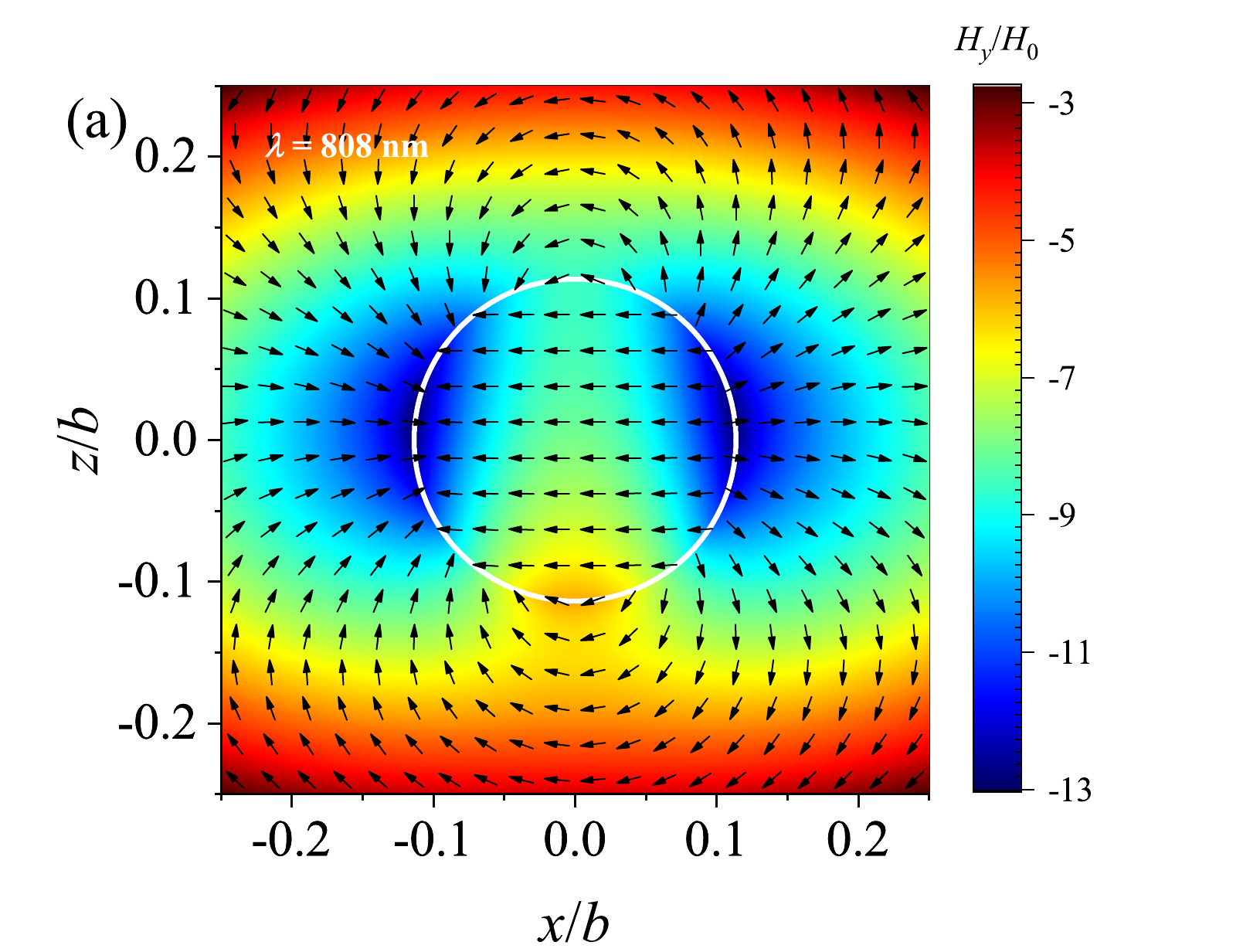}\\
\vspace{-0.5cm}
\includegraphics[width=0.9\columnwidth]{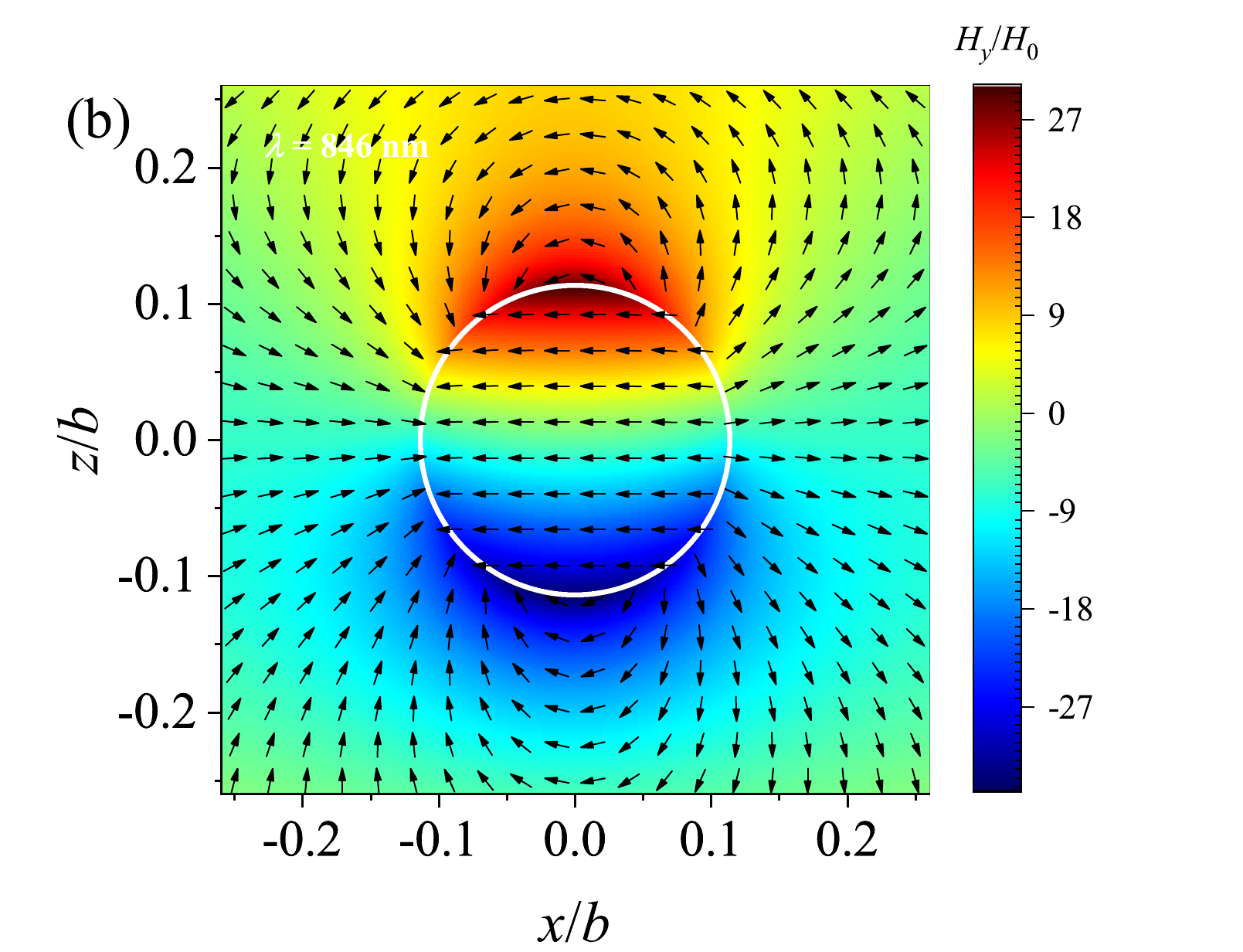}
\caption{Near-field distribution for an (Ag) core-shell (AlGaAs) nanosphere of inner radius $a=25$~nm and outer radius $b=220$~nm without gain ($\kappa=0$).
The distribution of $H_y$ (color plots) and $\mathbf{E}$ (vector plots) in $xz$ plane in the vicinity of the Ag core for (a) $\lambda=808$~nm and (b) $\lambda=846$~nm.
}\label{fig7}
\end{figure}

To exemplify a practical application of our analytic expressions, let us first consider an Ag nanosphere of radius $a=25$~nm coated with an AlGaAs spherical shell $(n_2=3.5-\imath\kappa)$ of radius $b=220$~nm (layer thickness $b-a=195$~nm) illuminated by a linear-polarized plane electromagnetic wave, as illustrated in Fig.~\ref{fig2}.
The plots of the scattering cross section and first-order contributions to light scattering as a function of the wavelength are presented in the left panel of Fig.~\ref{fig4}.
The electric dipole and toroidal dipole excitations are calculated using the analytic expressions derived for the partial scattering coefficients in Eqs.~(\ref{ae-analytic}) and (\ref{at-analytic}).
In the range of wavelengths from $790$ to $890$~nm, the size parameter varies into the interval $1.5<kb<1.75$, which is beyond the small-particle limit.
However, the dipole approximations derived for the partial scattering coefficients still holds for these values of size parameters in the dynamic Lorenz-Mie regime, \color{black} which can be verified via Eq.~(\ref{a1-approx-equal}).\color{black}

In Fig.~\ref{fig4}(a), we illustrate the effect of electric toroidal dipole-induced transparency in plasmonic core-shell spheres, as discussed in Ref.~\cite{Miroshnichenko_LaserPhotRev9_2015}.
The destructive interference between the partial coefficients $a_1^{\rm c}$ (c-ED, dotted line) and $a_1^{\rm t}$ (ETD, solid line) results in a highly suppressed $\sigma_{\rm sca}$ (dashed line) for $\lambda\approx 808$~nm.
Particularly, for $\lambda\approx840$~nm, the plots in Fig.~\ref{fig4}(a) indicate that light scattering is primarily due to an ETD excitation, as the c-ED is significantly suppressed at this wavelength.
In fact, the interference between these two first-order electric excitations gives rise to a Fano lineshape in the spherical electric dipole $|a_1|^2$ (s-ED, dashed-dotted line) in this wavelength regime.
This asymmetric resonance, resulting from the interference between two profiles with different origins, can be used to explain the effect of electric toroidal dipole-induced transparency, where the Fano dip corresponds to the transparency window in the spectra.
Additionally, by considering the presence of gain medium in the dielectric nanoshell, as shown in Fig.~\ref{fig4}(b) ($n_2=3.5-0.015\imath$) and Fig.\ref{fig4}(c) ($n_2=3.5-0.025\imath$), we demonstrate that the ETD contribution to light scattering can be amplified while maintaining the suppression of the c-ED excitation for $\lambda\approx 840$~nm and simultaneously preserving the toroidal dipole-induced transparency.

In the right panel of Fig.~\ref{fig4}, the normalized time-averaged electric and magnetic field intensities within both the plasmonic core ($|\mathbf{E}_1|^2$ and $|\mathbf{H}_1|^2$) and the gain-assisted dielectric shell ($|\mathbf{E}_2|^2$ and $|\mathbf{H}_2|^2$) are plotted.
These quantities are calculated by using the time-averaged intensities derived in Ref.~\cite{Arruda_JOpt14_2012}.
By comparing the plots in Figs.~\ref{fig4}(a) and \ref{fig4}(d), it can be observed that the ETD for $\lambda\approx840$~nm is mainly associated with the electric (solid line) and magnetic (dotted line) field resonances inside the plasmonic core.
The resonance peaks in the electric (dashed line) and magnetic (dash-dotted line) field intensities inside the dielectric shell are associated with the electric quadrupole excitation $(\ell=2)$.
In particular, for a gain-assisted dielectric shell, these resonances related to first-order excitations inside the core-shell sphere can be amplified as the gain coefficient is increased.
This can be seen in Fig.~\ref{fig4}(e) for $n_2=3.5-0.015\imath$ and Fig.~\ref{fig4}(f) for $n_2=3.5-0.025\imath$.
Interestingly, in Fig.~\ref{fig4}(f), it is shown that $n_2=3.5-0.025\imath$ also leads to the suppression of the field intensities associated with the electric quadrupole resonances at $\lambda\approx857$~nm in the gain-assisted dielectric shell, while the field intensities inside the plasmonic core are enhanced.

To further clarity the multipole interference that leads to the suppression of $\sigma_{\rm sca}$, in Fig.~\ref{fig5} we plot the real and imaginary parts of the partial coefficients associated with the Cartesian electric dipole and electric toroidal dipole.
From these plots, it is clear that the effect of electric toroidal dipole-induced transparency occurs for $\lambda\approx 808$~nm when $a_1^{\rm c}\approx - a_1^{\rm t}$.
In particular, as shown in Fig.~\ref{fig4}(b), the maximum electric and magnetic field intensities inside the plasmonic core are achieved for $\lambda\approx 846$~nm.
At this wavelength, we observe in Fig.~\ref{fig5} that the real part of $a_1^{\rm t}$ assumes its maximum value, with ${\rm Re}(a_1^{\rm t})\approx{\rm Im}(a_1^{\rm t})\approx{\rm Re}(a_1^{\rm c})$, and ${\rm Im}(a_1^{\rm c})\approx0$.
The resonance in the electric and magnetic field intensities at the same wavelength inside the plasmonic core are associated with an interplay between the electric dipole and electric toroidal dipole excited within the core-shell particle.

To analyze the electric and magnetic fields inside the (Ag) core-shell (AlGaAs) sphere, in Fig.~\ref{fig6} we use the full-wave Lorenz-Mie theory to calculate  the electric $(E_x)$ and magnetic $(H_y)$ near-field distributions perpendicular to a plane containing the $z$-axis and passing through the center of the spherical particle.
The normalized electric vector field (black arrows) in $xz$ plane is shown in Fig.~\ref{fig6}(a), while the normalized magnetic vector field (black arrows) in $yz$ plane is depicted in Fig.~\ref{fig6}(b).
As already verified in the left panel of Fig.~\ref{fig4}, a gain-assisted nanoshell affects the strength of field intensities inside the scatterer, but not their resonance wavelength.
Hence, for the sake of simplicity, we choose a dielectric shell without gain ($\kappa=0$) to study the near-field distribution, which corresponds to the configuration presented in Figs.~\ref{fig4}(a) and \ref{fig4}(d).

The field components $H_y$ (magnetic field along the $y$-direction) and $E_x$ (electric field along the $x$-direction) are calculated in Figs.~\ref{fig6}(a) and \ref{fig6}(b), respectively, for $\lambda=808$~nm, which is approximately the wavelength associated with the electric toroidal dipole-induced transparency $(\sigma_{\rm sca}\approx 0)$.
As one can observe in Figs.~\ref{fig6}(a) and \ref{fig6}(b), both the magnetic and electric near-field distributions oscillate outside the core-shell sphere with $|H_y|\approx H_0$ and $|E_x|\approx E_0$, which are the electric and magnetic field amplitudes of the incident electromagnetic plane wave.
However, inside the core-shell sphere, we note a nontrivial field distribution for $H_y$ in Fig.~\ref{fig6}(a) in $xz$ plane, with a significant magnetic field amplitude inside the plasmonic core and dielectric shell but with opposite signs.
As for Fig.~\ref{fig6}(b) in $yz$ plane, we see that the electric field $E_x$ is highly concentrated within the plasmonic core.
By analyzing the electric vector-field pattern in Fig.~\ref{fig6}(a), we see the presence of two vortices in the dielectric shell around the maximum amplitude of the magnetic field component $H_y$.
In the vicinity of the Ag core, we observe a near-field distribution associated with the ED excitation.
Indeed, this a typical configuration of a non-radiating charge-current configuration, also known as a dynamic anapole: the ED excited within the Ag core is canceled out by the ETD excited within the dielectric shell.

Figures~\ref{fig6}(c) and \ref{fig6}(d) depict the same configuration of an (Ag) core-shell (AlGaAs) sphere as discussed above, but for $\lambda=846$~nm.
This wavelength corresponds to the resonance wavelength of the magnetic field intensity inside the plasmonic core calculated in Fig.~\ref{fig4}(d).
The presence of higher-order excitations is clearly seen in the $H_y$ near-field distribution shown in Fig.~\ref{fig6}(c) and in the magnetic field vortices exhibited in Fig.~\ref{fig6}(d).
Although these higher-order excitations appear in the near-field distribution, one can observe in Fig.~\ref{fig6}(c) a typical near-field distribution pattern of the ETD inside the plasmonic core, with the approximate oscillation of $H_y/H_0$ between $-27$ to $27$.
In fact, the main contribution within the core is due to the transverse-magnetic field components of the first order, with small contribution of radial magnetic field components.

To better visualize what is happening within the core-shell sphere, Figs.~\ref{fig7}(a) and \ref{fig7}(b) show the near-field distributions in the vicinity of the Ag core for $\lambda= 808$~nm and $846$~nm, respectively, and $\kappa=0$ as in Figs.~\ref{fig6}(a) and \ref{fig6}(c).
In Fig.~\ref{fig7}(a), we clearly observe a dipole scattering pattern around the plasmonic core, which interferes destructively with the electric toroidal dipole excited within the dielectric nanoshell, see Fig.~\ref{fig6}(a).
The change in the direction of the normalized electric vector field within the Ag core when compared with its vicinity is due to the negative permittivity exhibited by the metallic materials, and can be explained by the electrostatic induction in a subwavelength metallic structure $(ka\ll1)$.
Figure~\ref{fig7}(b) shows that the resonance of the internal field intensities inside the plasmonic core for $\lambda\approx 846$~nm is related to the excitation of the electric toroidal dipole in the vicinity of the Ag core, with the characteristic circulation of the electric field around the maxima of magnetic field forming a torus shape, see Fig.~\ref{fig6}(d).

\subsection{Magnetic toroidal dipole in a dielectric sphere coated with a plasmonic nanoshell}
\label{Ag-shell}

\begin{figure*}[htbp]
\hspace{-.9cm}
\includegraphics[width=.45\textwidth]{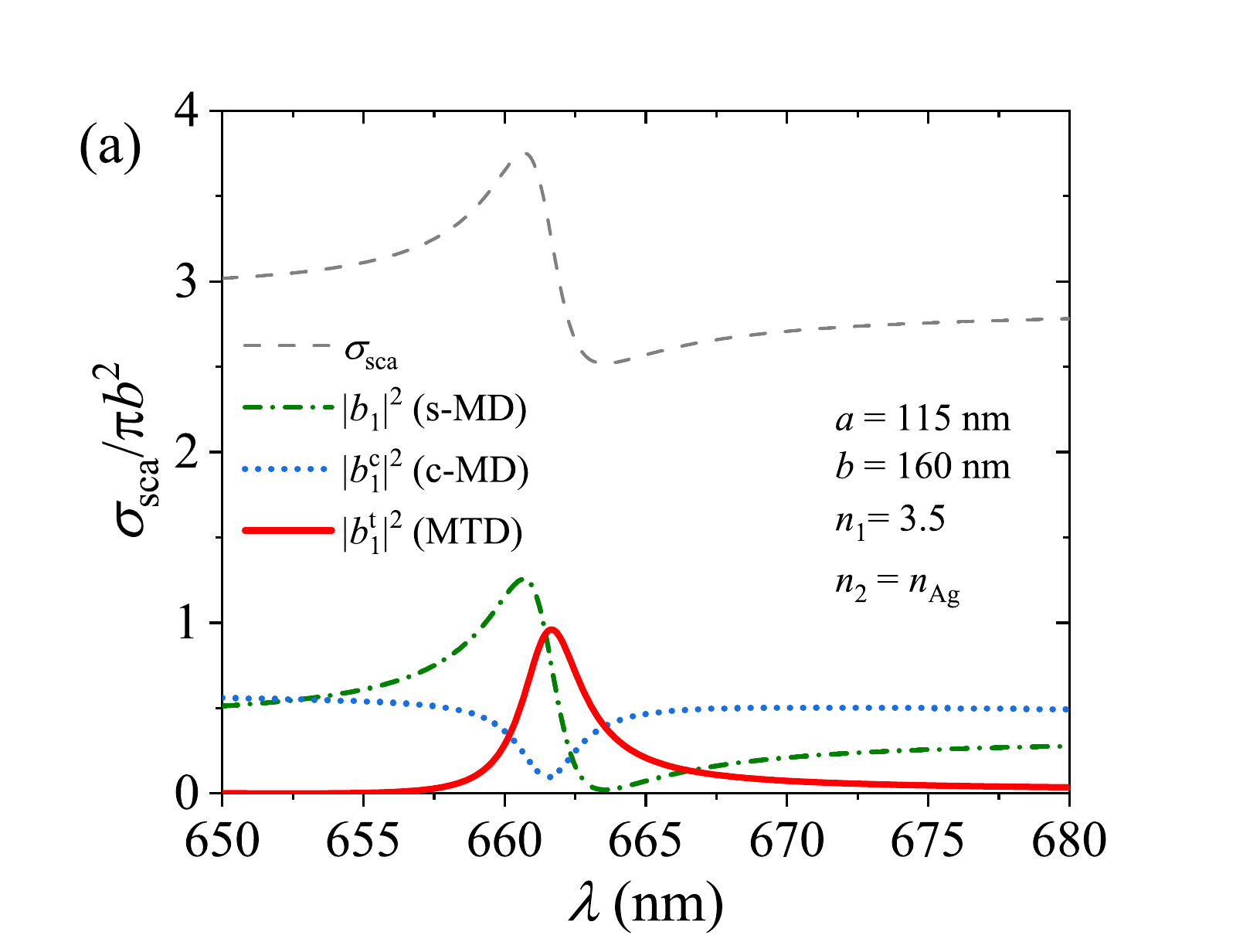}
\includegraphics[width=.45\textwidth]{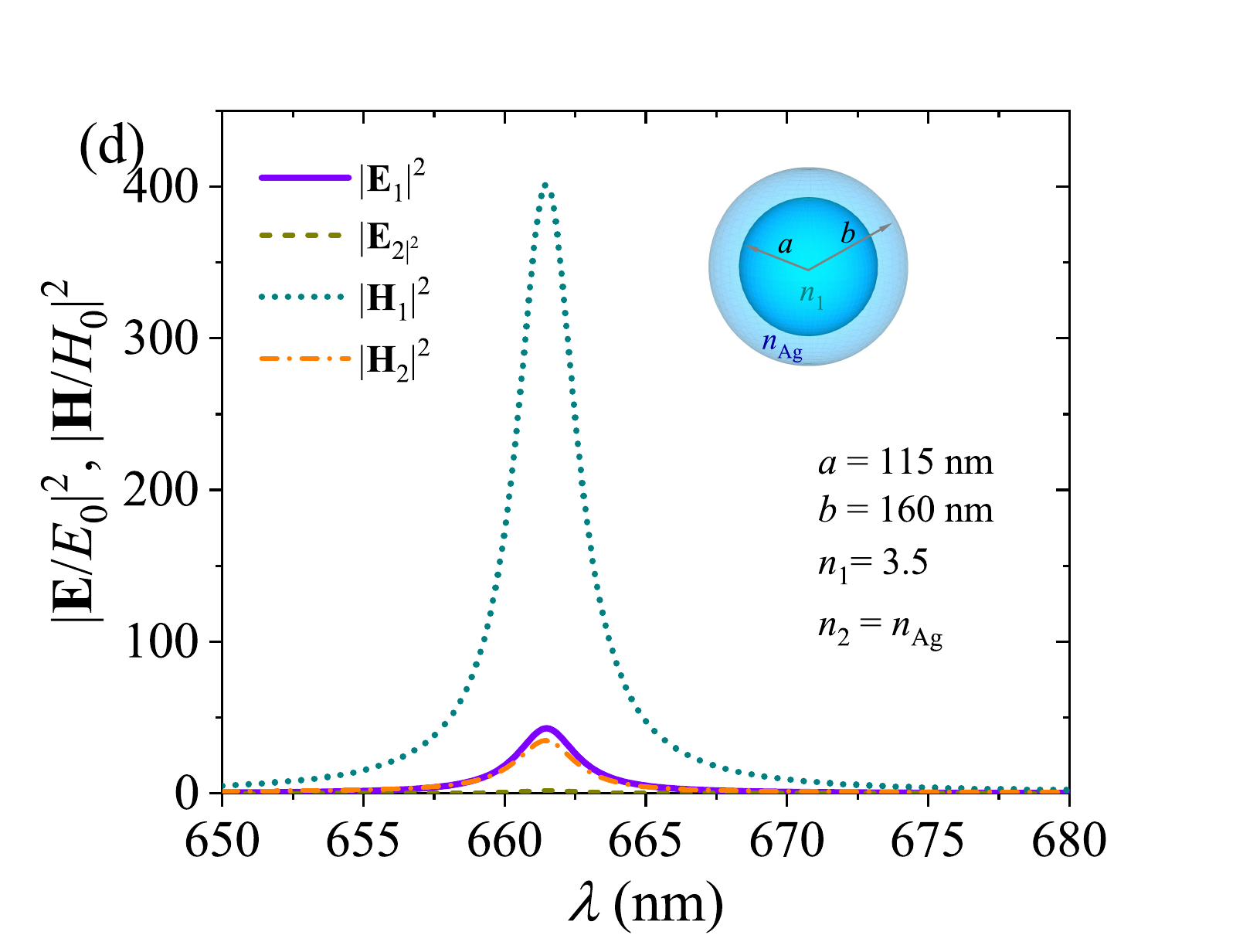}\\
\vspace{-.5cm}
\hspace{-.9cm}
\includegraphics[width=.45\textwidth]{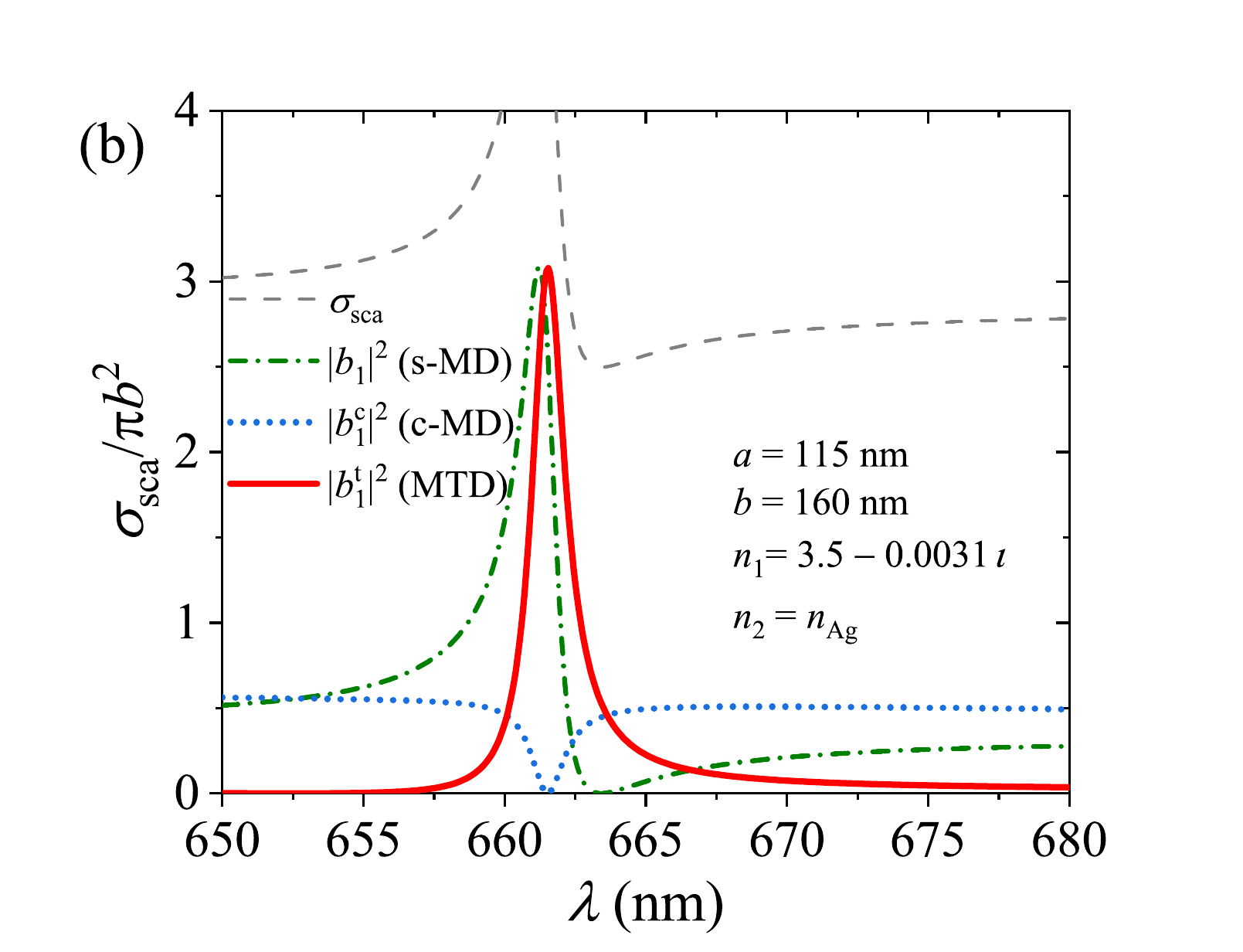}
\includegraphics[width=.45\textwidth]{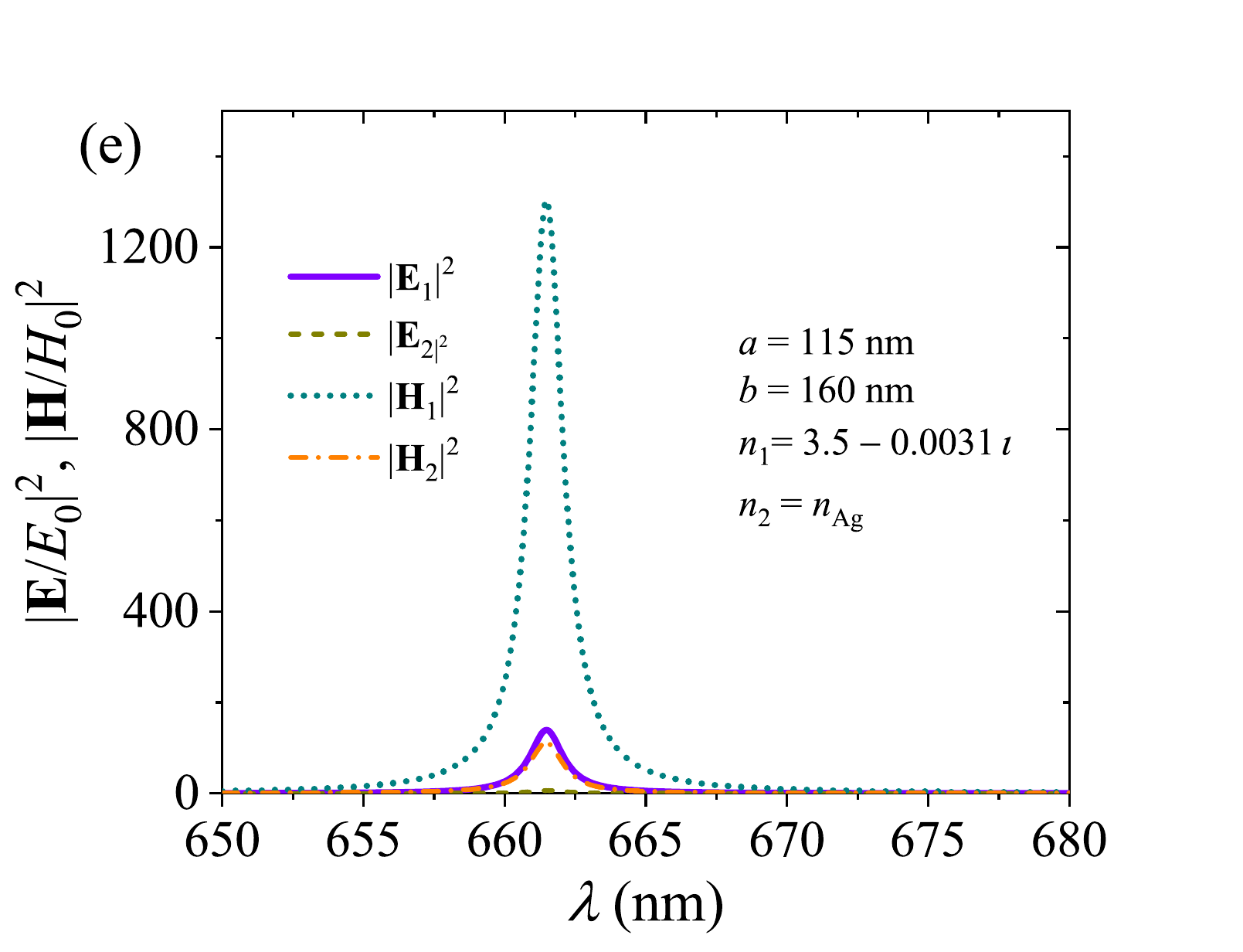}\\
\vspace{-.5cm}
\hspace{-.9cm}
\includegraphics[width=.45\textwidth]{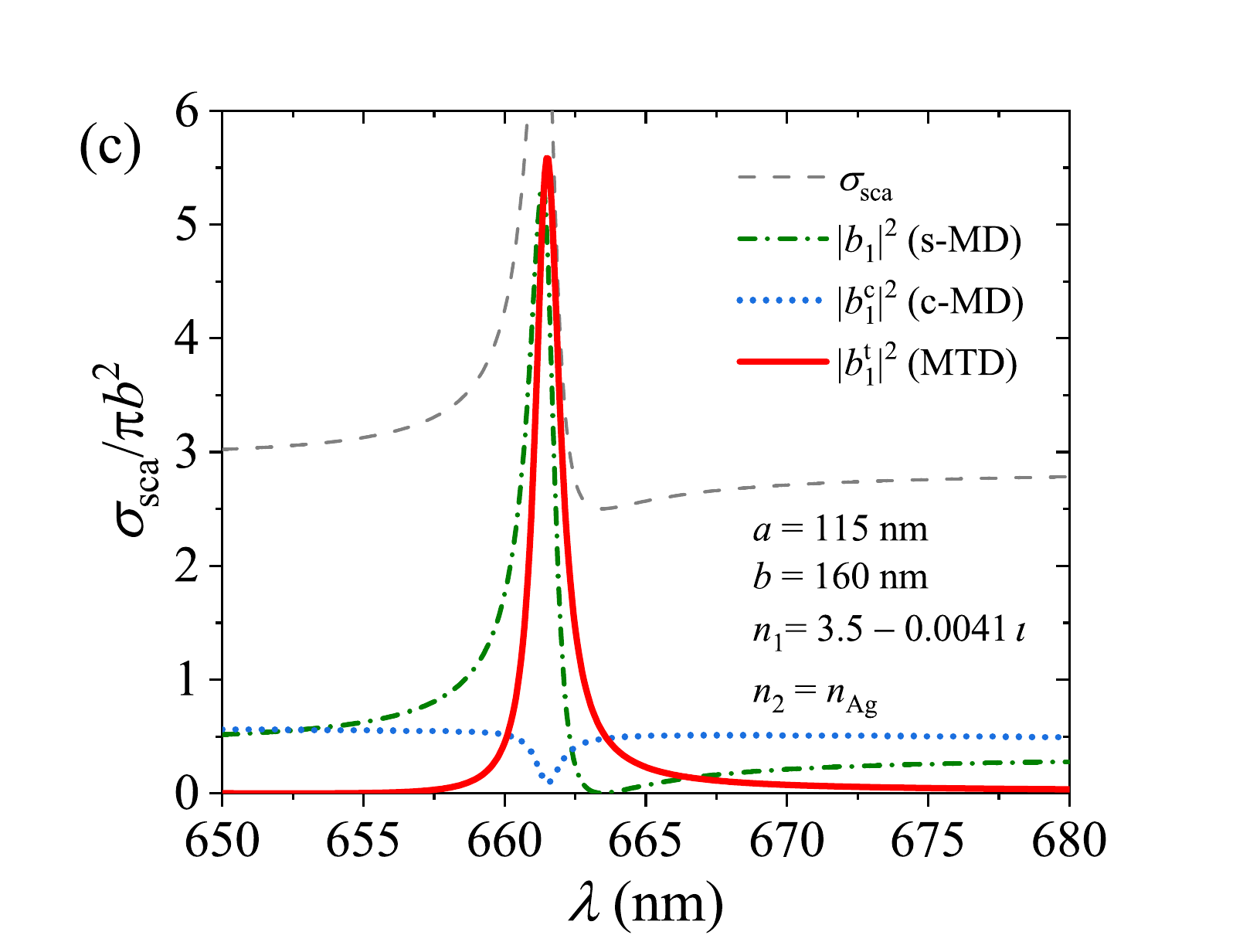}
\includegraphics[width=.45\textwidth]{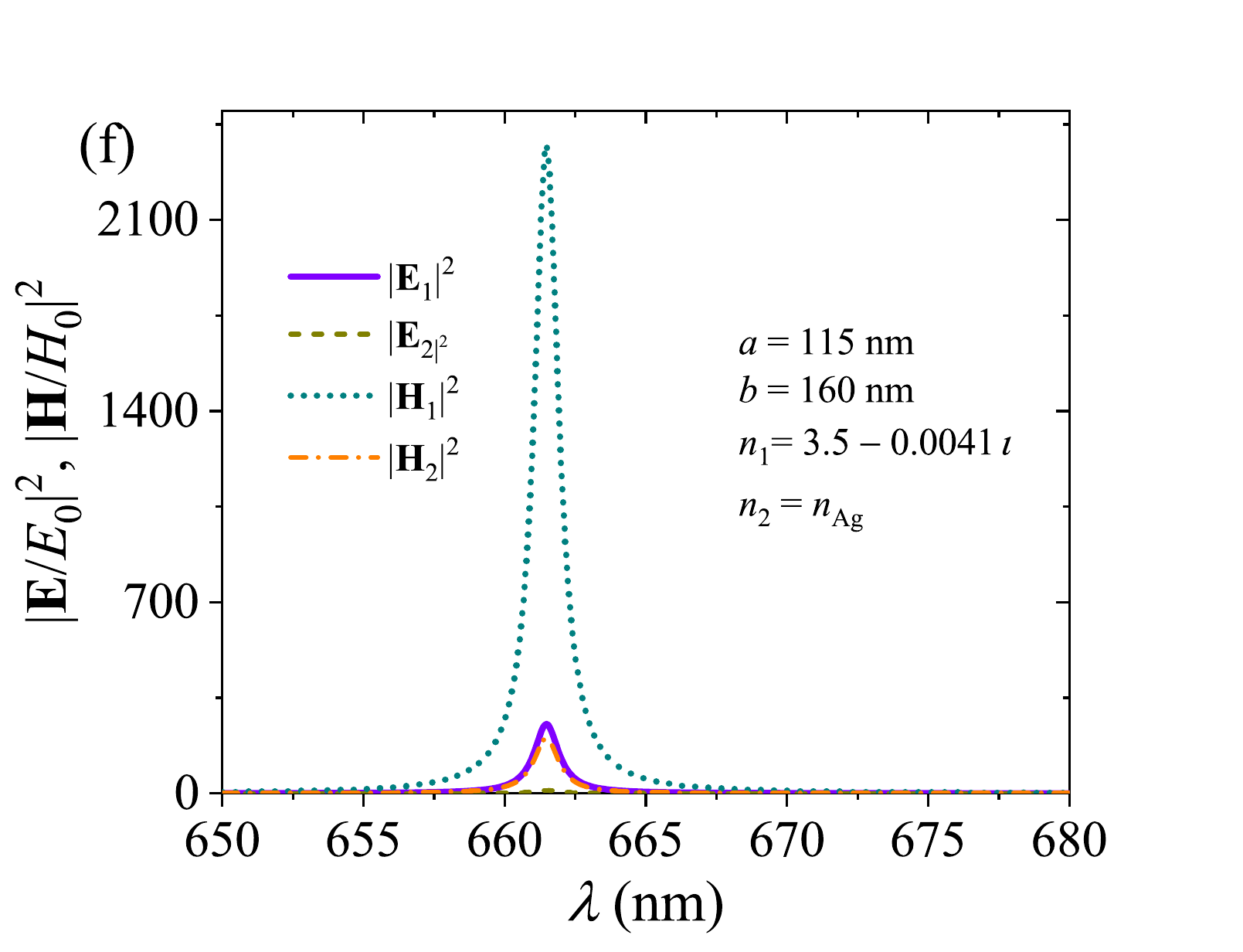}
\caption{An AlGaAs nanosphere $(n_1=3.5-\imath \kappa)$ of radius $a=115$~nm coated with an Ag nanoshell of radius $b=160$~nm, with $\kappa$ a gain coefficient. The scattering cross-section $\sigma_{\rm sca}$ (dashed line) and the contributions of the spherical magnetic dipole associated with $b_1$ (s-MD, dashed-dotted-dotted line), the Cartesian magnetic dipole with $b_1^{\rm c}$ (c-MD, dotted line), and the magnetic toroidal dipole with $b_1^{\rm t}$ (MTD, solid line) are plotted as a function of the wavelength for (a) $\kappa=0$, (b) $\kappa=0.0031$, and (c) $\kappa=0.0041$. The corresponding electric $|\mathbf{E}_q|^2$ and magnetic $|\mathbf{H}_q|^2$ field intensities inside the plasmonic core ($q=1$) and dielectric shell $(q=2)$ are also shown for (d) $\kappa=0$, (e) $\kappa=0.0031$, and (f) $\kappa=0.0041$.}
\label{fig8}
\end{figure*}

\begin{figure}[htbp!]
\includegraphics[width=0.9\columnwidth]{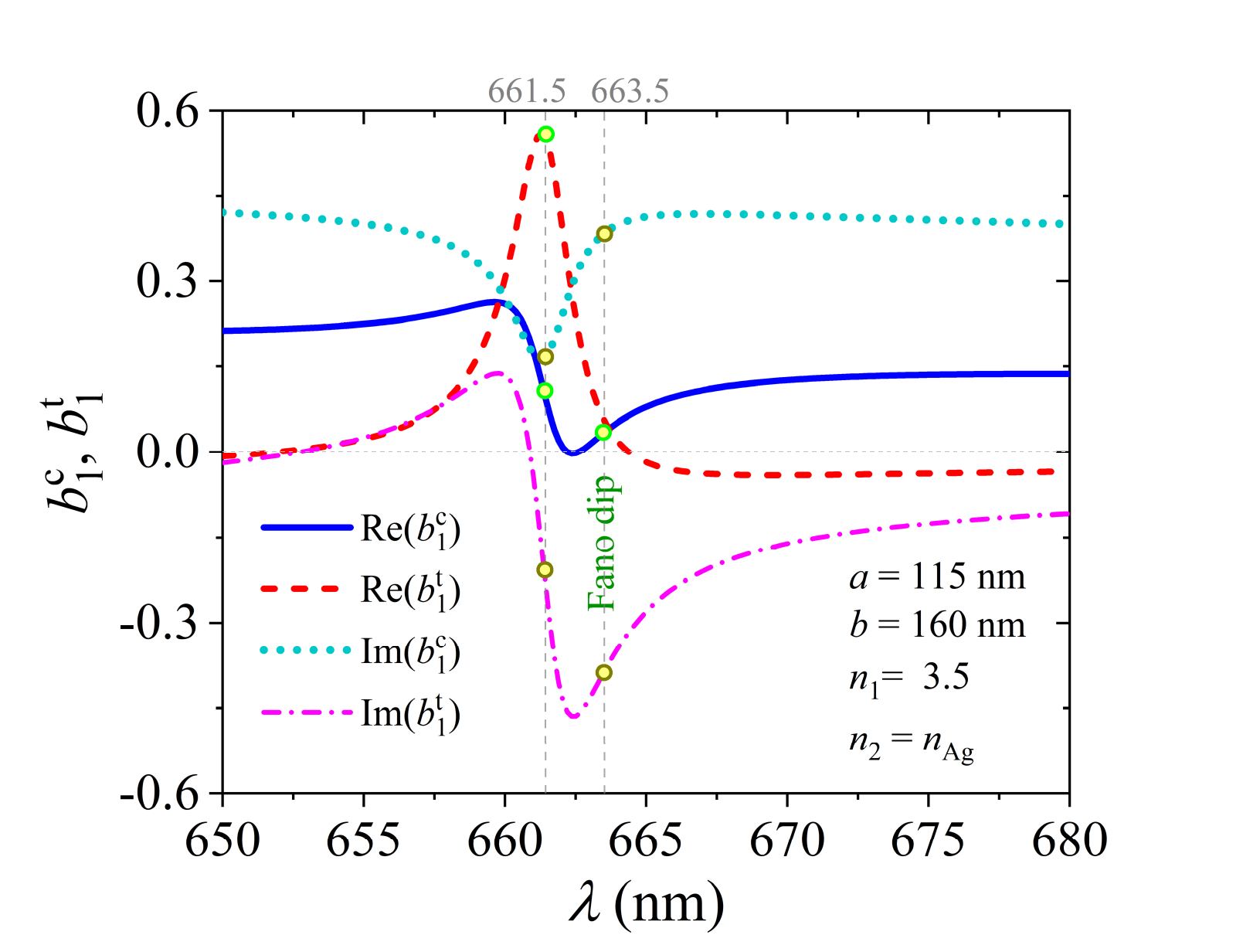}
\caption{First-order scattering coefficients associated with light scattering by a dielectric sphere ($n_1=3.5$) of radius $a=115$~nm coated with an Ag nanoshell of radius $b=160$~nm.
The plots show the real and imaginary parts of the partial coefficients $b_1^{\rm c}$ and $b_1^{\rm t}$, which are related to the Cartesian magnetic dipole and magnetic toroidal dipole, respectively.
}\label{fig9}
\end{figure}

\begin{figure*}[htbp]
\hspace{-.9cm}
\includegraphics[width=.45\textwidth]{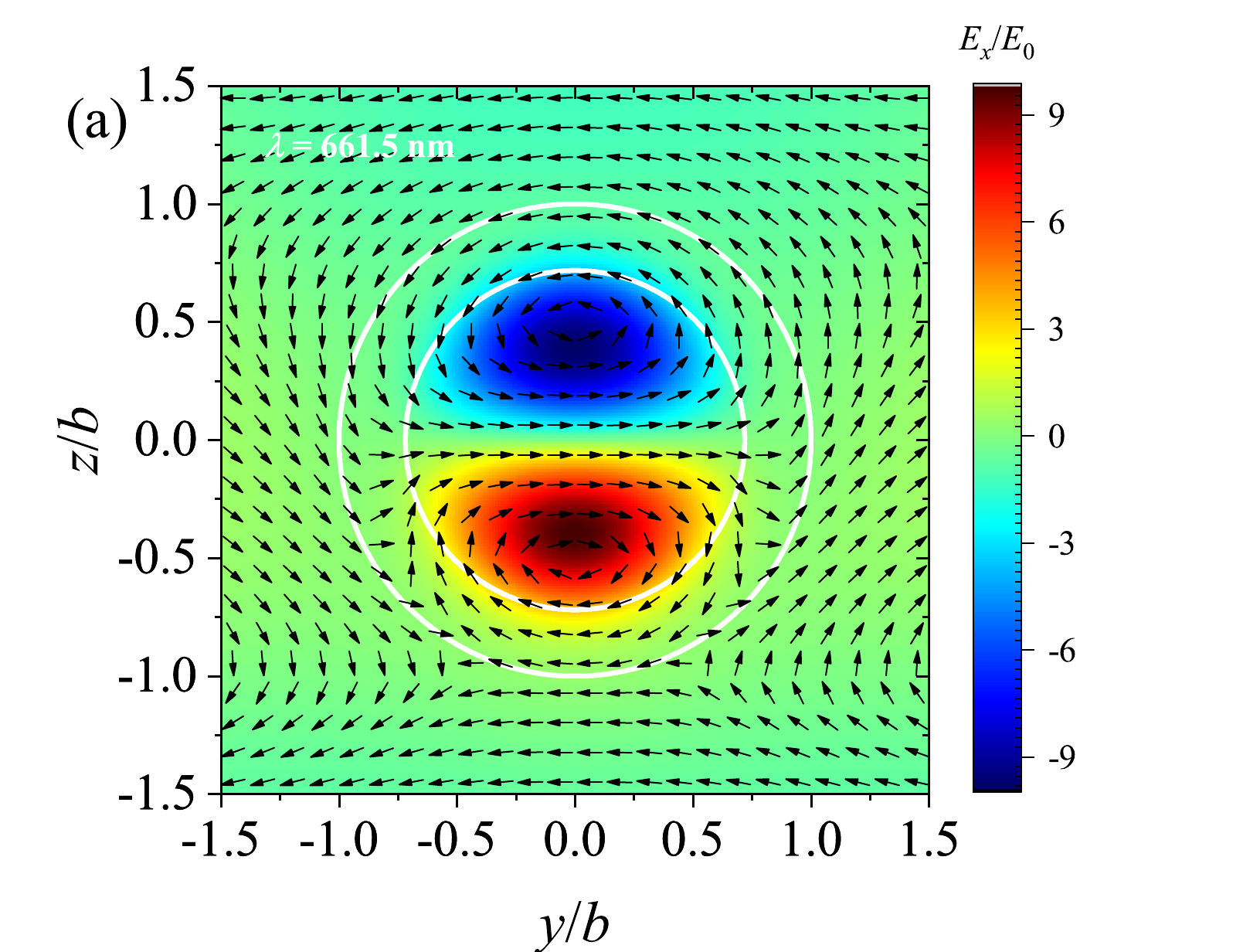}\hspace{-1cm}
\includegraphics[width=.45\textwidth]{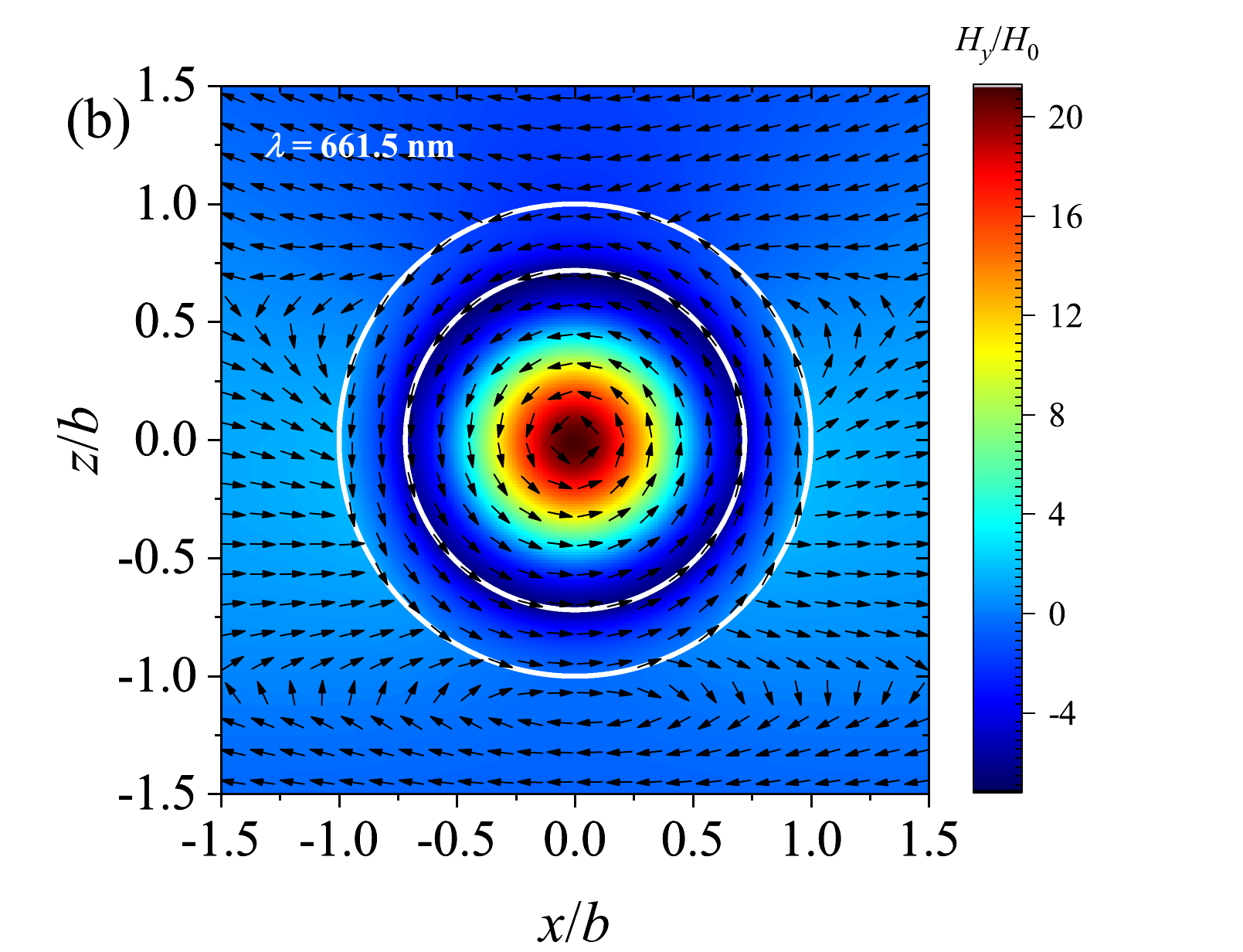}\\
\vspace{-.5cm}
\hspace{-.9cm}
\includegraphics[width=.45\textwidth]{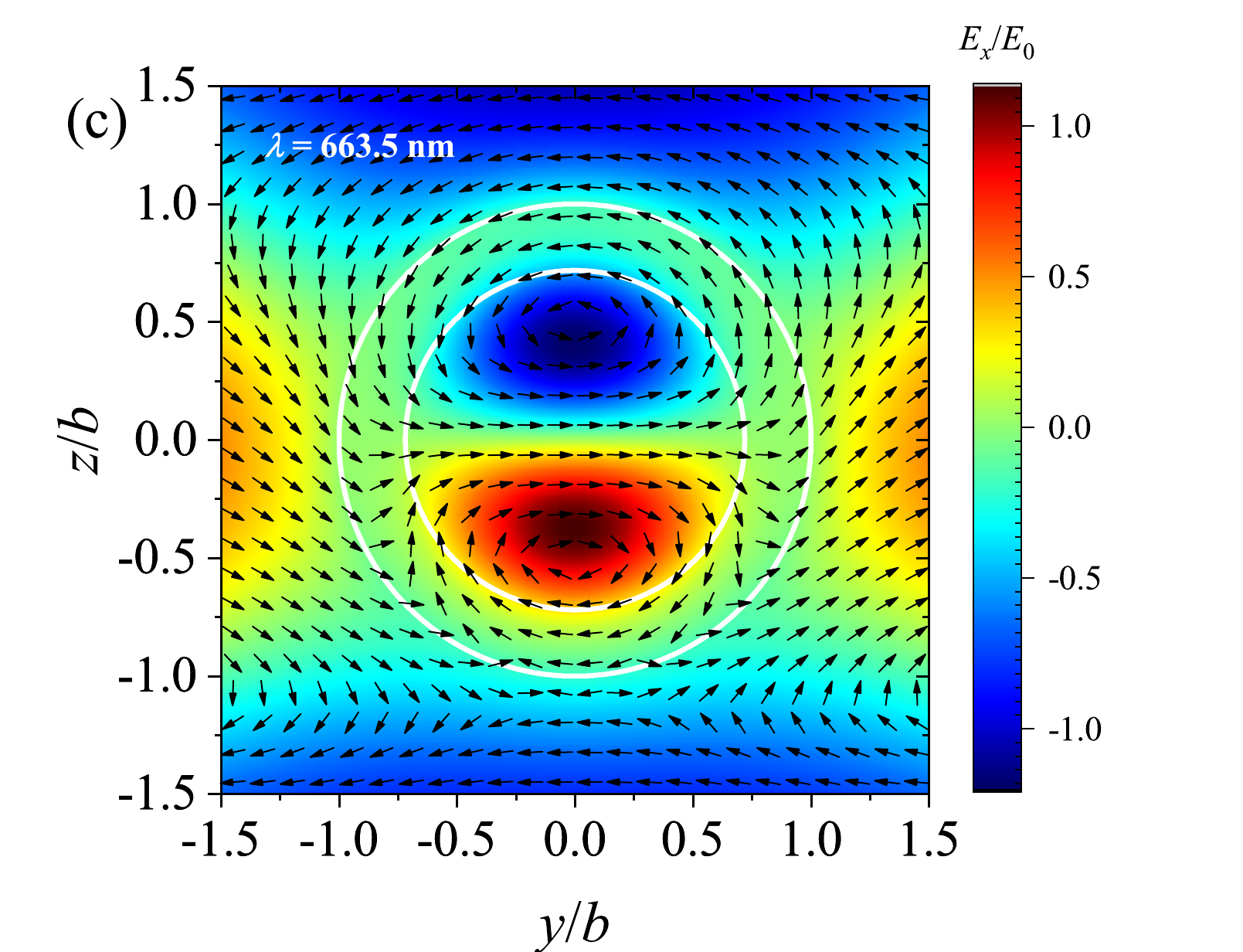}\hspace{-1cm}
\includegraphics[width=.45\textwidth]{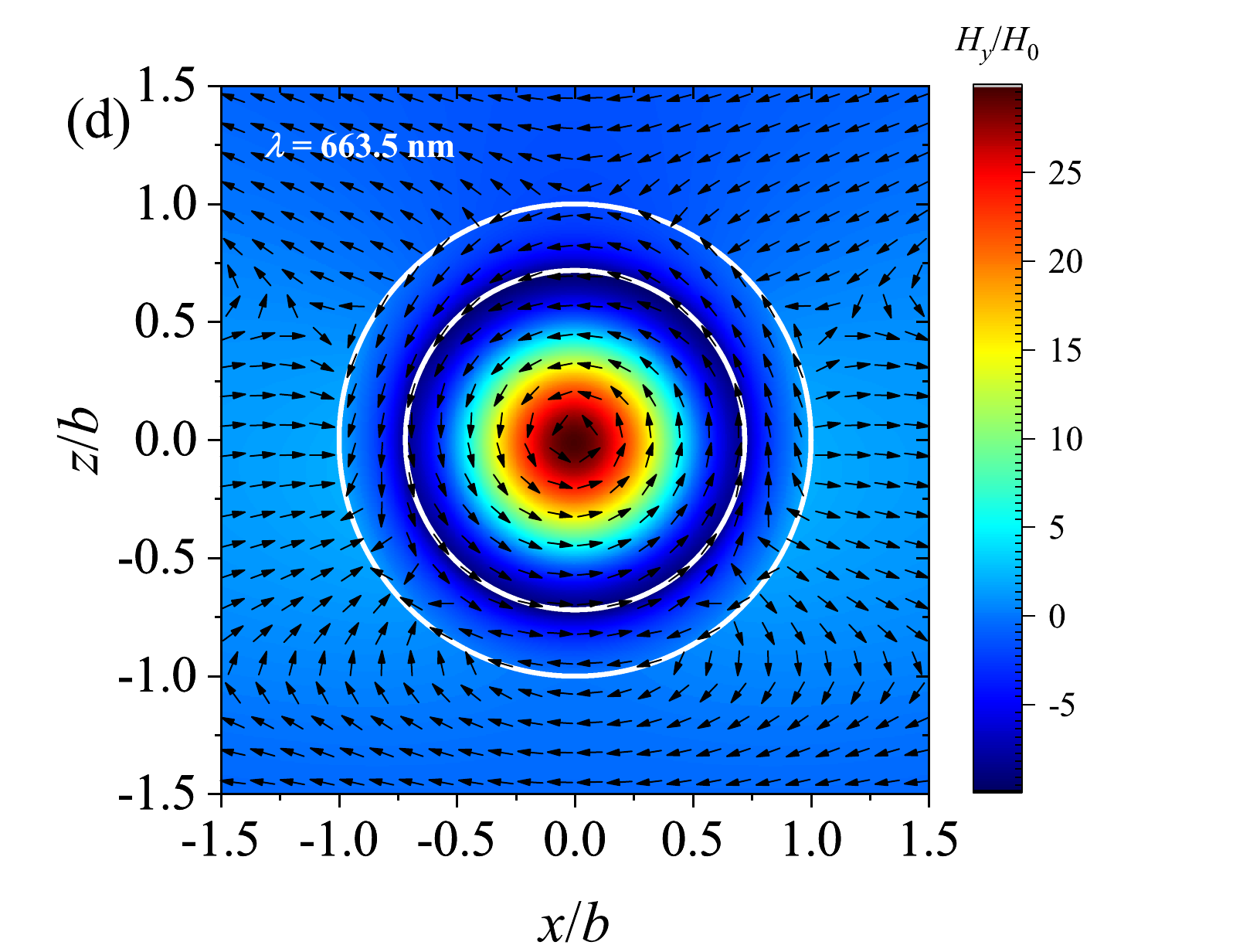}
\caption{Electric and magnetic field distributions in a plane across the center of a dielectric $(n_1=3.5)$ sphere of radius $a=115$~nm coated with an Ag nanoshell of radius $b=160$~nm.  (a) The distribution of both ${E}_x$ (color plots) and $\mathbf{H}$ (vector plots) in the $yz$ plane for $\lambda=661.5$~nm, which corresponds to the resonance peak in the magnetic field intensity in Fig.~\ref{fig8}(b).
(b) The distribution of both ${H}_y$ (color plots) and $\mathbf{E}$ (vector plots) in the $xz$ plane for $\lambda=661.5$~nm.
(c) The distribution of both ${E}_x$ (color plots) and $\mathbf{H}$ (vector plots) in the $yz$ plane for $\lambda=663.5$~nm, which corresponds to the Fano dip in the renormalized magnetic dipole in Fig.~\ref{fig8}(a).
(d) The distribution of both ${H}_y$ (color plots) and $\mathbf{E}$ (vector plots) in the $xz$ plane for $\lambda=663.5$~nm.
}\label{fig10}
\end{figure*}

In this section we consider the light scattering by a dielectric AlGaAs sphere ($n_1=3.5-\imath\kappa$) of radius $a=115$~nm coated with an Ag nanoshell of radius $b=160$~nm in free space.
Our aim is to excite the c-MD and MTD in the Lorenz-Mie scattering for $kb\approx 1$.
To this end, we consider linearly-polarized plane electromagnetic waves with wavelengths in the range of $650$ to $680$~nm, which corresponds to the size parameter interval of $1.47<kb<1.55$.
In this range, higher-order multipoles $(\ell>1)$ are excited within the plasmonic core-shell nanosphere.
\color{black}
However, the dipole approximation given by Eq.~(\ref{b1-approx-equal}) still holds for the partial magnetic coefficients.
\color{black}

The left panel of Fig.~\ref{fig8} presents plots of the scattering cross section and first-order contributions to light scattering by a (AlGaAs) core-shell (Ag) nanosphere for different values of the gain coefficient $\kappa$ as a function of wavelength.
It is clearly shown in Fig.~\ref{fig8}(a) that the first-order magnetic excitations are not sufficient to describe the light scattering.
However, one can verify that other contributions, which must be taken into accounted in the total scattering cross section $\sigma_{\rm sca}$, such as the conventional spherical electric dipole $|a_1|^2$ (s-ED) and the conventional electric quadrupole $|a_2|^2$, remain approximately constant for the chosen parameters.
Therefore, the profile shown in $\sigma_{\rm sca}$ (dashed line) as a function of wavelength is mainly due to the spherical magnetic dipole $|b_1|^2$ (s-MD, dash-dotted line) as it can be shifted vertically to match the total scattering cross section.

In Fig.~\ref{fig8}(a), one can observe that the Cartesian magnetic dipole associated with $b_1^{\rm c}$ (c-MD, dotted line) is suppressed at $\lambda\approx 663$~nm.
This dip in the c-MD contribution to light scattering coincides with a peak in the Cartesian magnetic toroidal dipole associated with $b_1^{\rm t}$ (MTD, solid line).
Indeed, the interference between these two first-order contributions gives rise to a Fano lineshape, which is shown in the s-MD associated with the $b_1$.
Since there are higher-order electric excitations contributing to light scattering in this wavelength regime, the Fano dip in the s-MD does not lead to $\sigma_{\rm sca}\approx 0$ (dashed line).
However, since the other contributions are approximately constant within this wavelength regime, the resonance peak in $\sigma_{\rm sca}$ is clearly due to Fano resonance in the s-MD.

We show in Fig.~\ref{fig8}(d) that the Fano resonance in the scattering cross section ($\lambda\approx 660.6$~nm) is associated with an off-resonance field enhancement inside the dielectric core ($\lambda\approx 661.5$~nm).
The resonance in the internal field intensities matches the wavelength in which the Cartesian magnetic dipole related to $b_1^{\rm c}$ presents a dip in the spectra ($\lambda\approx 661.5$~nm), which approximately coincides with the resonance in the MTD associated with $b_1^{\rm t}$ ($\lambda\approx661.7$).
Hence, the off-resonance field enhancement inside an (AlGaAs) core-shell (Ag) nanosphere can be explained by a compromise between near-field excitations ($b_1^{\rm t}$, MTD) and far-field excitations ($b_1^{\rm c}$, c-MD), whose interference leads to a Fano lineshape.
This result, obtained from $b_1^{\rm c}$ and $b_1^{\rm t}$, is hidden in the Lorenz-Mie coefficient $b_1$ associated with the s-MD, and cannot be directly retrieved from $\sigma_{\rm sca}$.

The left panel of Fig.~\ref{fig8} demonstrates the possibility of enhancing the MTD contribution to light scattering, while maintaining the suppression of the c-ED by adding gain to the AlGaAs core.
In Figs.~\ref{fig8}(b) and \ref{fig8}(c), the strength of the resonance peaks associated with MTD is approximately multiplied by three (for $n_1=3.5-0.0031\imath$) and by six (for $n=3.5-0.0041\imath$), respectively, when compared to the case without gain in Fig.~\ref{fig8}(a).
Additionally, we observe in Figs.~\ref{fig8}(e) and \ref{fig8}(f) that the gain medium leads to a significant enhancement of the magnetic field intensity $|\mathbf{H}_1|^2$ within the AlGaAs core in the same proportion as the MTD in the light scattering spectra.
This implies that one can control the amount of the electromagnetic energy inside a core-shell scatterer by enhancing the MTD contribution to $\sigma_{\rm sca}$, which can be done by using a gain-assisted dielectric sphere coated with a plasmonic nanoshell.
We emphasize that the superscattering effect achieved in Fig.~\ref{fig8}(c) around $663$~nm is owing mainly owing to the MTD, as the s-ED and higher-order contributions remain unchanged by the gain addition, and the c-MD is almost suppressed.

The results obtained in Figs.~\ref{fig8}(a) and \ref{fig8}(d) for an (AlGaAs) core-shell (Ag) sphere without gain can be explained via the partial coefficients $b_1^{\rm c}$ and $b_1^{\rm t}$.
For $\lambda=661.5$~nm, Fig.~\ref{fig9} shows that ${\rm Re}(b_1^{\rm t})$ is maximum whereas ${\rm Im}(b_1^{\rm c})$ is minimum, which coincides with the resonance in the MTD in $\sigma_{\rm sca}$.
In addition, for $\lambda=663.5$~nm, there is a Fano dip in the s-MD contribution, which corresponds to the condition $b_1^{\rm c}=-b_1^{\rm t}$.
In this case, as mentioned above, the suppression of the s-MD does not lead to $\sigma_{\rm sca}\approx 0$ (transparency) due to the contribution of the s-ED and electric quadrupole in the spectrum.

In Fig.~\ref{fig10}, we plot the near-field distributions of field components $E_x$ (electric field along the $x$ direction) and $H_y$ (magnetic field along the $y$ direction) in a perpendicular plane passing through the center of the (AlGaAs) core-shell (Ag) nanosphere for $\kappa=0$ (no gain).
Figure~\ref{fig10}(a) shows the $E_x$ distribution in the $xy$ plane for $\lambda=661.5$~nm, which approximately corresponds to the resonance wavelength of the MTD and the suppression of the c-MD.
The vector plot (black arrows) corresponds to the normalized magnetic vector field in the $xy$ plane.
Conversely, in Fig.~\ref{fig10}(b), we show the $H_y$ distribution in the $xz$ plane for the same wavelength, where the vector plot (black arrows) depicts the normalized electric vector field in $xz$.
As anticipated, since only the MTD is resonant at this frequency, Figs.~\ref{fig10}(a) and \ref{fig10}(b) show a near-field distribution pattern typical of a pure magnetic toroidal dipole excitation, with the circulation of the magnetic vector field around the maxima of the electric field in Fig.~\ref{fig10}(a) leading to a cross section of a torus in Fig.~\ref{fig10}(b).
Additionally, in Figs.~\ref{fig10}(c) and \ref{fig10}(d), we plot the same near-field distribution for $\lambda=663.5$~nm, which is associated with the Fano dip exhibited in $\sigma_{\rm sca}$.
Although there are higher-order excitations that modify the near-field distribution at this wavelength, we still obtain the magnetic toroidal pattern in the near-field distribution.
However, the far-field pattern in Fig.~\ref{fig10}(c) is shown to be dominated by the s-ED as the s-MD is suppressed.

We emphasize that the results presented in this section and in Sec.~\ref{Ag-core}, for both cases of a plasmonic core-shell nanosphere containing a linear-gain material, are not optimized for specific applications.
Our aim is simply to illustrate the possibility of exciting both electric and magnetic toroidal dipole moments in light scattering by core-shell nanospheres using the Lorenz-Mie theory.
Importantly, we demonstrate how one could use the new scattering coefficients along with the near-field distribution to obtain a more comprehensive understanding of multipole interferences, leading to phenomena such as Fano resonances, transparency, and superscattering of light.
Other intriguing possibilities for manipulating toroidal multipole interactions remain open for further investigation, employing the novel scattering coefficients in diverse core-shell configurations of materials and parameters.

\section{Conclusion}
\label{conclusion}

In conclusion, we have derived novel scattering coefficients within the framework of the Lorenz-Mie theory concerning electric and magnetic toroidal dipoles.
These new coefficients, previously concealed within the spherical electric and magnetic dipole contributions, accurately replicate the results obtained from the Cartesian electric and magnetic toroidal dipoles.
As an application of these analytic expressions, we have investigated light scattering by a plasmonic core-shell sphere, a fundamental component of metamaterials, in various configurations incorporating a gain-assisted dielectric medium in the scatterer.
For an Ag nanosphere coated with an AlGaAs nanoshell, we have demonstrated the feasibility of probing electric toroidal dipole-induced transparency using the newly calculated scattering coefficients, where the transparency condition is simply stated as $a_1^{\rm c}=-a_1^{\rm t}$.
The possibility of enhancing the electric toroidal dipole contribution via a gain-assisted dielectric shell was also discussed.
Furthermore, for an AlGaAs nanosphere coated with an Ag nanoshell, we have illustrated the potential for exciting magnetic toroidal dipole resonances in the far field while suppressing the Cartesian magnetic dipole.
By employing a gain-assisted dielectric core, we have achieved a superscattering configuration attributed solely to the magnetic toroidal dipole response.
We believe that the investigation of electric and magnetic toroidal dipole excitations in light scattering by spherical particles will greatly benefit from the novel scattering coefficients $a_1^{\rm c}$, $a_1^{\rm t}$, $b_1^{\rm c}$ and $b_1^{\rm t}$, thus opening new avenues for exploring scattering phenomena involving interference of multipoles in the Lorenz-Mie theory.

\section{Acknowledgment}

The author thanks Felipe A. Pinheiro for fruitful discussions at an early stage of this study.

\appendix

\section{Lorenz-Mie scattering coefficients}\label{an-and-bn}

The electric and magnetic Lorenz-Mie scattering coefficients associated with a core-shell sphere are calculated from boundary conditions~\cite{Bohren_Book_1983}.
These coefficients describe the interaction of an incident plane electromagnetic wave with the scatterer, and are given by~\cite{Arruda_JOpt14_2012}
\begin{align}
        a_{\ell} &=\frac{n_2\psi_{\ell}'(kb)-\psi_{\ell}(kb)\mathcal{A}_{\ell}(n_2kb)}{n_2\xi_{\ell}'(kb)-\xi_{\ell}(kb)\mathcal{A}_{\ell}(n_2kb)}\
        ,\label{an}\\
        b_{\ell} &=\frac{\psi_{\ell}'(kb)-n_2\psi_{\ell}(kb)\mathcal{B}_{\ell}(n_2kb)}{\xi_{\ell}'(kb)-n_2\xi_{\ell}(kb)\mathcal{B}_{\ell}(n_2kb)}\
        ,\label{bn}
\end{align}
where the auxiliary functions are
\begin{align}
&\mathcal{A}_{\ell}(n_2kb)=\frac{\psi_{\ell}'(n_2kb)-A_{\ell}\chi_{\ell}'(n_2kb)}{\psi_{\ell}(n_2kb)-A_{\ell}\chi_{\ell}(n_2kb)},\label{A3}\\
&\mathcal{B}_{\ell}(n_2kb)=\frac{\psi_{\ell}'(n_2kb)-B_{\ell}\chi_{\ell}'(n_2kb)}{\psi_{\ell}(n_2kb)-B_{\ell}\chi_{\ell}(n_2kb)},\\
&        A_{\ell} = \frac{n_2\psi_{\ell}(n_2ka)\psi_{\ell}'(n_1ka)-n_1\psi_{\ell}'(n_2ka)\psi_{\ell}(n_1ka)}{n_2\chi_{\ell}(n_2ka)\psi_{\ell}'(n_1ka)-n_1\chi_{\ell}'(n_2ka)\psi_{\ell}(n_1ka)},\\
&        B_{\ell}=\frac{n_2\psi_{\ell}'(n_2ka)\psi_{\ell}(n_1ka)-n_1\psi_{\ell}(n_2ka)\psi_{\ell}'(n_1ka)}{n_2\chi_{\ell}'(n_2ka)\psi_{\ell}(n_1ka)-n_1\chi_{\ell}(n_2ka)\psi_{\ell}'(n_1ka)}.\label{A6}
\end{align}

If the charge-current distribution $\mathbf{J}(\mathbf{r})$ within the scatterer is known, one can calculate these coefficients using Eqs.~(\ref{a_E2}) and (\ref{a_M2}).
In fact, by substituting the expressions of the internal electric fields, Eqs.~(\ref{E1}) and (\ref{E2}), into Eq.~(\ref{J}), and plugging $\mathbf{J}(\mathbf{r})$ in Eqs.~(\ref{a_E2}) and (\ref{a_M2}), we obtain after some algebra:

\begin{align}
a_{\ell}&=\imath\left(1-n_1^2\right)d_{\ell}\int_0^{ka}{\rm d}x\psi_{\ell}(x)\frac{\psi_{\ell}(n_1x)}{n_1^2}\nonumber\\
&+\imath\left(1-n_2^2\right)g_{\ell}\int_{ka}^{kb}{\rm d}x\psi_{\ell}(x)\frac{\left[\psi_{\ell}(n_2x)-A_{\ell}\chi_{\ell}(n_2x)\right]}{n_2^2}\nonumber\\
&+\imath\left(1-n_1^2\right)d_{\ell}\int_0^{ka}{\rm d}x\left[\psi_{\ell}'(x)\frac{\psi_{\ell}'(n_1x)}{n_1}+\psi_{\ell}''(x)\frac{\psi_{\ell}(n_1x)}{n_1^2}\right]\nonumber\\
&+\imath\left(1-n_2^2\right)g_{\ell}\int_{ka}^{kb}{\rm d}x\bigg\{\psi_{\ell}'(x)\frac{\left[\psi_{\ell}'(n_2x)-A_{\ell}\chi_{\ell}'(n_2x)\right]}{n_2}\nonumber\\
&+ \psi_{\ell}''(x)\frac{\left[\psi_{\ell}(n_2x)-A_{\ell}\chi_{\ell}(n_2x)\right]}{n_2^2}\bigg\},\label{ae-int}\\
b_{\ell}&=\imath\left(1-n_1^2\right)c_{\ell}\int_0^{ka}{\rm d}x\psi_{\ell}(x)\frac{\psi_{\ell}(n_1x)}{n_1}\nonumber\\
&+\imath\left(1-n_2^2\right)f_{\ell}\int_{ka}^{kb}{\rm d}x\psi_{\ell}(x)\frac{\left[\psi_{\ell}(n_2x)-B_{\ell}\chi_{\ell}(n_2x)\right]}{n_2},\label{b-int}
\end{align}
where $c_{\ell}$, $d_{\ell}$, $f_{\ell}$ and $g_{\ell}$ are the Lorenz-Mie coefficients of the fields within the core-shell sphere, Eqs.~(\ref{cn})-(\ref{gn}).
To simplify the angular integrals, we have used the relations~\cite{Bohren_Book_1983}:
\begin{align}
&\int_{-1}^{1}{\rm d}(\cos\theta)\pi_{\ell1}\pi_{\ell'1}\sin^2\theta=\frac{2\ell(\ell+1)}{(2\ell+1)}\delta_{\ell\ell'},\label{pi1}\\
&\int_{-1}^1{\rm d}(\cos\theta)(\pi_{\ell1}\pi_{\ell'1}+\tau_{\ell1}\tau_{\ell'1})=\frac{2\ell^2(\ell+1)^2}{(2\ell+1)}\delta_{\ell\ell'},\\
&\int_{-1}^1{\rm d}(\cos\theta)(\pi_{\ell1}\tau_{\ell'1}+\tau_{\ell}\pi_{\ell'1})=0,\label{pi3}
\end{align}
where $\delta_{\ell\ell'}$ is the Kronecker's delta.

The two last integrals in Eq.~(\ref{ae-int}) can be simplified via integration by parts:
\begin{align}
&\int_0^{ka}{\rm d}x\left[\psi_{\ell}'(x)\frac{\psi_{\ell}'(n_1x)}{n_1}+\psi_{\ell}''(x)\frac{\psi_{\ell}(n_1x)}{n_1^2}\right]\nonumber\\
&=\psi_{\ell}'(x)\frac{\psi_{\ell}(n_1x)}{n_1^2}\bigg|_{x=0}^{x=ka},\\
&\int_{ka}^{kb}{\rm d}x\bigg\{\psi_{\ell}'(x)\frac{\left[\psi_{\ell}'(n_2x)-A_{\ell}\chi_{\ell}'(n_2x)\right]}{n_2}\nonumber\\
&+ \psi_{\ell}''(x)\frac{\left[\psi_{\ell}(n_2x)-A_{\ell}\chi_{\ell}(n_2x)\right]}{n_2^2}\bigg\}\nonumber\\
&=\psi_{\ell}'(x)\frac{\left[\psi_{\ell}(n_2x)-A_{\ell}\chi_{\ell}(n_2x)\right]}{n_2^2}\bigg|_{x=ka}^{x=kb}\label{ae-var}.
\end{align}
The remaining integrals involving Bessel functions can be calculated by the relation
\begin{align}
\int_{l_1}^{l_2} {\rm d}x \psi_{\ell}(x)\varrho(nx)&=\frac{n\psi_{\ell}(x)\varrho_{\ell}'(nx)-\psi_{\ell}'(x)\varrho_{\ell}(nx)}{1-n^2}\bigg|_{x=l_1}^{x=l_2}
\end{align}
where $\varrho(nx)$ can be either $\psi_{\ell}(nx)$ or $\chi_{\ell}(nx)$~\cite{Abramovitz_1964,Arruda_JOSA27_2010,Arruda_JOpt14_2012}.
Finally, to simplify the final expressions, we have to consider the boundary condition equations
$\psi_{\ell}(n_1ka) d_{\ell}= g_{\ell}[\psi_{\ell}(n_2ka)-A_{\ell}\chi_{\ell}(n_2ka)]$ and $\psi_{\ell}(n_1ka) c_{\ell}=f_{\ell}[\psi_{\ell}'(n_2ka)-B_{\ell}\chi_{\ell}'(n_2ka)]$~\cite{Bohren_Book_1983}.
Using all of these relations, we can readily solve the integrals in Eqs.~(\ref{ae-int}) and (\ref{b-int}), leading to

\begin{align}
a_{\ell}&=
-\imath \frac{g_{\ell}}{n_2^2}\bigg\{\psi_{\ell}'(kb)\left[\psi_{\ell}(n_2kb)-A_{\ell}\chi_{\ell}(n_2kb)\right]\nonumber\\
&-n_2\psi_{\ell}(kb)\left[\psi_{\ell}'(n_2kb)-A_{\ell}\chi_{\ell}'(n_2kb)\right]\bigg\}\nonumber\\
&-\imath\left({1}-\frac{1}{n_2^2}\right)g_{\ell}\psi_{\ell}'(kb)\left[\psi_{\ell}(n_2kb)-A_{\ell}\chi_{\ell}(n_2kb)\right],\label{ae-ell}\\
b_{\ell}
&=-\imath \frac{f_{\ell}}{n_2}\bigg\{\psi_{\ell}'(kb)\left[\psi_{\ell}(n_2kb)-B_{\ell}\chi_{\ell}(n_2kb)\right]\nonumber\\
& -n_2\psi_{\ell}(kb)\left[\psi_{\ell}'(n_2kb)-B_{\ell}\chi_{\ell}'(n_2kb)\right]\bigg\}.\label{b-ell}
\end{align}

Finally, using the exact expressions for the internal coefficients $f_{\ell}$ and $g_{\ell}$, provided in Eqs.~(\ref{fn}) and (\ref{gn}), one can easily retrieve Eqs.~(\ref{an}) and (\ref{bn}).

\end{document}